\documentclass[a4paper,11pt]{article}
\pdfoutput=1 

\usepackage{jheppub} 

\usepackage[T1]{fontenc} 

\usepackage{amsfonts}
\usepackage{amsmath}
\usepackage{amssymb}
\usepackage{multirow}

\usepackage{caption}
\DeclareMathOperator*{\IM}{Im}
\DeclareMathOperator*{\res}{res}

\usepackage{graphicx}

\title{Rigorous Bounds on Light-by-Light Scattering}

\author[\gamma]{J. Henriksson,}
\author[\gamma]{B. McPeak,}
\author[\gamma]{F. Russo,}
\author[\gamma]{A. Vichi,}

\affiliation[\gamma]{Department of Physics, University of Pisa and INFN, \\Largo Pontecorvo 3, I-56127 Pisa, Italy}

\emailAdd{johan.henriksson@df.unipi.it}
\emailAdd{brian.mcpeak@df.unipi.it}
\emailAdd{francesco.russo@phd.unipi.it}
\emailAdd{alessandro.vichi@unipi.it}

\abstract{We bound EFT coefficients appearing in $2 \to 2$ photon scattering amplitudes in four dimensions. After reviewing unitarity and positivity conditions in this context, we use dispersion relations and crossing symmetry to compute sum rules and null constraints. This allows us to derive new rigorous bounds on operators with four, six, and eight derivatives, including two-sided bounds on their ratios. Comparing with a number of partial UV completions, we find that some of our bounds are saturated by the amplitudes that arise from integrating out a massive scalar or axion, while others suggest the existence of unknown amplitudes. }

\begin{document} 
\maketitle
\flushbottom

\section{Introduction}

Many physical systems display multiple characteristic energy scales, where they are described by different degrees of freedom and governed by different dynamics. When these scales are well separated, it is possible to integrate out the high-energy excitations and restrict attention to low-energy ones, which will be described by a new set of interactions.
The resulting Effective Field Theory (EFT) represents a powerful framework to parametrize the ignorance about the microscopic behavior of a model. Instead of worrying about the detailed ultraviolet (UV) completion of a theory, one can instead specify the low-energy content of the theory and expand the interaction in a controlled series of higher-dimension operators made with a few building blocks and suppressed by increasing powers of the microscopic scale. The coefficients of this expansion, also called Wilson coefficients, encode the UV details of the system.

It was realized some time ago \cite{Pham:1985cr,Ananthanarayan:1994hf,Adams:2006sv} that not all values of these coefficients are consistent with a well behaved UV completion: 
despite the spirit of the EFT approach to be as agnostic as possible about the microscopic description of the theory, there are essential properties we should not abandon if we want the underlying theory to be consistent with basic axioms obeyed by $S$-matrices. Enforcing these conditions on the EFT produces a set of linear inequalities involving the Wilson coefficients.

More recently, new approaches  \cite{Arkani-Hamed:2020blm,Tolley:2020gtv,Bellazzini:2020cot,Caron-Huot:2020cmc,Sinha:2020win} have exploited more stringent assumptions on the UV theory to extract two-sided bounds on ratios of Wilson coefficients. 
In a sense, these results put the usual dimensional analysis on rigorous ground, fixing the strength of EFT coefficients to an $O(1)$ number times the appropriate inverse power of the cutoff scale.

The mentioned works rely on basic properties of scattering amplitudes, such as unitarity, causality, crossing symmetry, existence of a partial wave decomposition and the behavior of the amplitude at infinity, to convert dispersion relations into sum rules involving the EFT data. 
In order to interpret these constraints in terms of bounds on the Wilson coefficients, it is crucial to restrict to weakly coupled EFT up to the cutoff scale, so that one can legitimately neglect loops of low-energy degrees of freedom. 
On one hand this restriction limits the space of EFTs to which these bounds apply; on the other hand, it allows a simpler derivation of the results, combining numerical and analytic techniques. This approach may be thought of as an ``EFT bootstrap''.

Complementary to the above works, the $S$-matrix bootstrap \cite{Paulos:2016fap,Paulos:2016but,Paulos:2017fhb} has developed systematic methods to construct the most general scattering amplitude consistent with basic axioms of quantum field theory. 
By scanning over all possible amplitudes, one can explore the allowed values of observables such as interactions, masses of resonances, etc. 
Recent applications were also able to fix the low-energy behavior of an amplitude in order to reproduce a given EFT \cite{Guerrieri:2020bto,Guerrieri:2021ivu}, while allowing the most general UV behavior.

The EFT bootstrap shares with the $S$-matrix bootstrap the goal of constraining the space of consistent quantum field theories. These programs have deep similarities with the conformal bootstrap program \cite{Rattazzi:2008pe}, both in their general philosophy and in their specific methods -- most notably, in the use of semi-definite programming to efficiently find optimal bounds. The EFT bootstrap further benefits from the use of the EFT as an organizing principle, as the constraints are very simple when expressed in terms of EFT coefficients. The two-sided bounds obtained for EFT coefficients are reminiscent of the islands of allowed values for operator scaling dimensions and couplings in the conformal bootstrap. As we will discuss, the analogy extends even further: we find examples of known UV completions appearing at kinks on the boundary of the allowed region. This raises the exciting possibility that the EFT bootstrap could be an efficient method for finding unknown UV completions for a given set of low-energy fields. 

\subsection{Photons EFT and summary of results}

\vspace{1em}

In this paper we continue the exploration of EFT constraints by focusing on photons in four dimensions, \emph{i.e.}\ massless spin one vectors.  Neglecting gravity, the photon is the only massless field in the Standard Model of particle physics which is not confined at low energies. Hence the electromagnetic gauge field $A_\mu$ is the only propagating degree of freedom in the infrared (IR). 
An observer performing experiments at center of mass (COM) energies $\sqrt s$ much lower than mass of the lightest (charged) particle will only be able to scatter photons and the fundamental observables are therefore the scattering amplitudes $\mathcal A^{\lambda_1\lambda_2 \cdots\lambda_n}$, where $\lambda_i=\pm$ denotes the polarization of the $i-$th photon.

Such an observer might try to describe the outcome of the scattering experiments by an EFT with a Lagrangian of the form
\begin{align}\label{eq:Lagrwitha1a2intro}
    \mathcal{L} = -\frac{1}{4} F_{\mu \nu} F^{\mu \nu} + a_1 (F_{\mu \nu}F^{\mu \nu})^2 + a_2  (F_{\mu \nu}\tilde F^{\mu \nu})^2 + \ldots,
\end{align}
where $a_1$, $a_2$ are dimensionful constants (Wilson coefficients) and $\tilde F_{\mu\nu}=\frac12\epsilon_{\mu\nu\rho\sigma}F^{\rho\sigma}$.
Dimensional analysis implies that the importance of the Wilson coefficients grows with COM energy, meaning that our low-energy observer will conclude that ``new physics'' must appear at some energy scale $M$. We will refer to this new physics as a \emph{partial UV completion}, to emphasize that it may itself need further completion at some even higher energy scale.

In the Standard Model, the scale $M$ is determined by the electron mass $M^2=4m_e^2$, and the leading correction to the Wilson coefficients come from integrating out an electron loop, giving
\begin{equation}
\label{eq:QED}
a_1^{\mathrm{QED}}=\frac{4\alpha^2}{360m_e^4},\quad a_2^{\mathrm{QED}}=\frac{7\alpha^2}{360m_e^4}.
\end{equation}
With these values, \eqref{eq:Lagrwitha1a2intro} agrees with the famous Euler--Heisenberg Lagrangian, written down in the 1930's as a precursor to quantum electrodynamics (QED)
\cite{Euler:1936oxn,Euler:1935zz,Heisenberg:1935qt}.\footnote{
A proper time parametrization was developed within the QED framework in \cite{Schwinger:1951nm}. 
See also \cite{Dunne:2004nc} for a modern discussion.} At higher energies, the contributions from heavier charged particles of the Standard Model become important \cite{Jikia:1993tc, Gounaris:1998qk, Bern:2001dg}.

Additional contributions coming from physics beyond the Standard Model, such as a light axion, could alter these predictions, see section~\ref{sec:treelevel}. This has triggered several experimental efforts to measure deviations from \eqref{eq:QED}.
The Wilson coefficients in the Euler--Heisenberg Lagrangian may be observed in a magnetic birefringence experiment, where a strong applied magnetic field gives rise to an anisotropy in the optical index proportional to $a_2-a_1$.\footnote{
Restoring units, we have $n_\parallel-n_\perp=\frac{16\hbar^3}{c^5\mu_0}(a_2-a_1)B_{\mathrm{ext}}^2=3A_eB_{\mathrm{ext}}^2$ with $A_e=1.32\cdot 10^{-24}\, \mathrm T^{-2}$. At $2.5\, \mathrm T$, the measured birefringence is $(12\pm17)\cdot 10^{-23}$, which is not sufficient to distinguish the Euler--Heisenberg Lagrangian from the free photon $a_1=a_2=0$ \cite{Ejlli:2020yhk}.
} In addition, in a direct scattering experiment $\gamma\gamma\to\gamma\gamma$ of unpolarized beams, the total cross-section at leading order is proportional to $a_1^2+a_2^2-2/3 a_1 a_2$ (see \emph{e.g.}\ \cite{Costantini:1971cj}). 

More generally, we can ask what partial UV completions are allowed based on some very general assumptions such as unitarity, causality and analyticity of scattering amplitudes. 
This question has a long history in the case of a low-energy EFT of scalars \cite{Roy:1971tc,Pham:1985cr,Ananthanarayan:1994hf, Colangelo:2001df, Caprini:2003ta, Adams:2006sv, Manohar:2008tc}. The case of photons was investigated in \cite{Cheung:2014ega}, and massive vectors in \cite{Distler:2006if}.\footnote{Recently, the dispersive representation has also been applied to photon amplitudes in order to quantify hadronic light-by-light contributions to the muon anomalous magnetic moment, see \cite{Colangelo:2014dfa,Colangelo:2015ama,Colangelo:2017fiz}.}
These papers were able to derive positivity conditions on various Wilson coefficients, ``one-sided bounds''. In particular, \cite{Cheung:2014ega} showed $a_1+a_2\geqslant0$ using similar arguments involving dispersion relations in the forward limit, and $a_1\geqslant0$, $a_2\geqslant0$ individually using independent arguments involving the absence of tachyons and ghosts plus some stronger assumptions about the form of the UV completion. Later efforts \cite{Bellazzini2016talk, Falkowski, Bellazzini:2019xts} showed that both $a_1$ and $a_2$ are individually positive based on the dispersion relation arguments of \cite{Cheung:2014ega} alone. In this paper, we use the method of \cite{Caron-Huot:2020cmc,Caron-Huot:2021rmr} to show that it is possible to get two-sided bounds on the Wilson coefficients for low-energy electromagnetism. In particular, we show how crossing symmetry leads to redundancy in the low-energy description, allowing the derivation of \emph{null constraints}, which are non-trivial constraints on the high-energy partial wave densities. These drastically strengthen the possible bounds, especially when combined with semidefinite programming.

To give an idea of the various bounds we find, consider the following parametrizations of the low-energy amplitudes:\footnote{In our conventions, the helicity indices represent ingoing particles, \emph{i.e.} $\mathcal A^{++--}$ corresponds $\gamma_+\gamma_+\to \gamma_+\gamma_+$.}
\begin{align}
\begin{split}
\label{eq:gsintro}
&\mathcal A^{++--}_{\mathrm L}=g_2s^2+g_3s^3+g_{4,1}s^4+g_{4,2}s^2(s^2+t^2+u^2)+\ldots,\\
&\mathcal{A}_\text{L}^{++++} \ = \ f_2 (s^2 + t^2 + u^2) + f_3 stu + f_4 (s^2 + t^2 + u^2)^2 + \ldots \, ,
\end{split}
\end{align}
where the constants $g_2$ and $f_2$ are related to Euler--Heisenberg coefficients simply by
\begin{align}\label{eq:a1a2g2f2refintro}
    a_1 = \frac{g_2 + f_2}{16} \, , \qquad a_2 = \frac{g_2 - f_2}{16}.
\end{align}
In the present work, we find a number of further interesting bounds, including:
\begin{itemize}
\item A two-sided bound for the ratio $g_3/g_2$ of the form \mbox{$- 4.828427<\frac{g_3 M^2}{g_2} \leqslant1$}. The upper bound is optimal, while the lower bound is weakly sensitive to the numerical precision (number of null constraints), to be specified below.
\item A finite allowed region in the space spanned by the parameters $\frac{g_{4,1}M^4}{g_2}$, $\frac{g_{4,2}M^4}{g_2}$, see figure~\ref{fig:g41g42}. In this parameter space, we compare our bound on $(g_{4,1},g_{4,2})$ to the one found in \cite{Arkani-Hamed:2020blm}. 

\item A finite allowed region in the space of couplings $f_2,f_3,f_4, g_3,  g_{4,1}, g_{4,2}$, normalized to $g_2$ and multiplied by the appropriate power of $M$. Sections of this region are shown in figures~\ref{fig:f2-dependence}, \ref{fig:g3f3}, \ref{fig:g41g42}, \ref{fig:g3g4} and \ref{fig:g3f4}.
\end{itemize}
Perhaps more interestingly, we find that a number of our bounds are saturated by amplitudes that arise from integrating a single massive scalar or axion at tree-level. We find that this is not the case for amplitudes arising from integrating out a massive graviton. Thus the scalar and axion tree-level exchange amplitudes are extremal, and those of graviton exchange are not.

The outline of the paper is the following: In section~\ref{sec:setup}, we show explicitly how to parameterize the low-energy (EFT) amplitudes in terms of symmetric polynomials of Mandelstam invariants, and the high-energy amplitudes in terms of partial waves with unknown but positive spectral densities. 
In section~\ref{sec:three}, we derive sum rules and null constraints by applying dispersion relations to these amplitudes, and use use them in section~\ref{sec:results} to derive both analytic and numerical bounds on the EFT coefficients.
Section~\ref{sec:results} also contains a comparison with partial UV completions derived from integrating out massive particles at tree-level or one-loop level. We finish with a discussion and some appendices providing more details on some aspects mentioned in the main text.

\section{Set-Up}
\label{sec:setup}

The method we use in this paper can be summarized in three steps:
\begin{enumerate}
    \item \label{step:one} parametrize the low-energy amplitude as EFT coefficients times Mandelstam invariants,
    \item \label{step:two} parametrize the high-energy amplitude by partial waves,
    \item use dispersion relations to relate the low-energy and high-energy amplitudes.
\end{enumerate}
In this section, we explain how we perform the first two parts of this strategy.

\subsection{Four-photon amplitudes}

We would like to understand: what effective field theories of photons are consistent with unitarity, Lorentz invariant $S$-matrices in the ultraviolet? To begin to answer this question, we will consider scattering amplitudes with four external photons. Such amplitudes may be written using a basis of 16 different observables, $\mathcal{A}^{\lambda_1 \lambda_2 \lambda_3 \lambda_4}$, where $\lambda_i = \pm$ is the helicity of particle $i$, and we use all ingoing conventions. 

These amplitudes are related by parity, time-reversal, and boson exchange according to:
\begin{align}
    \mathcal{P}:& \qquad \mathcal{A}^{\lambda_1 \lambda_2 \lambda_3 \lambda_4} = \mathcal{A}^{- \lambda_1 - \lambda_2 -\lambda_3 -\lambda_4}, \\
    \mathcal{T}:& \qquad \mathcal{A}^{\lambda_1 \lambda_2 \lambda_3 \lambda_4} = \mathcal{A}^{-\lambda_3 -\lambda_4 -\lambda_1 -\lambda_2 }, \\
    \mathcal{B}:& \qquad \mathcal{A}^{\lambda_1 \lambda_2 \lambda_3 \lambda_4} = \mathcal{A}^{\lambda_2 \lambda_1 \lambda_4 \lambda_3}.
\end{align}
In what follows, we will assume that our theory satisfies all three of these symmetries. Together, $\mathcal{P}$, $\mathcal{T}$, and $\mathcal{B}$ reduce the original basis of 16 amplitudes to only five independent ones. We may further consider crossing symmetry, which leads to the following requirements:
\begin{align}
\begin{split}
    \mathcal{A}^{++++}(s, t, u)&: \qquad \text{fully $s$-$t$-$u$ symmetric}, \\
    \mathcal{A}^{++--}(s, t, u) &: \qquad \text{$t$-$u$ symmetric}, \\
    \mathcal{A}^{+--+}(s, t, u) &: \qquad = \mathcal{A}^{++--}(u, t, s), \\
    \mathcal{A}^{+-+-}(s, t, u) &: \qquad  =\mathcal{A}^{++--}(t, s, u), \\
    \mathcal{A}^{+++-}(s, t, u)&: \qquad \text{fully $s$-$t$-$u$ symmetric},
\end{split}
\end{align}
where we have used $s = -(p_1 + p_2)^2$, $t = -(p_1 - p_4)^2$, and $u = -(p_1 - p_3)^2$ satisfying $s+t+u=0$. Crossing symmetry therefore allows us to write the amplitudes in terms of only three functions:
\begin{align}
\label{eq:lowenergyparam}
\begin{split}
    \mathcal{A}_\text{L}^{++++} \ &= \ f(s, t, u), \\
    \mathcal{A}_\text{L}^{++--} \ &= \ g(s| t, u), \\
    \mathcal{A}_\text{L}^{+--+} \ &= \ g(t| s, u), \\
    \mathcal{A}_\text{L}^{+-+-} \ &= \ g(u| t, s), \\
    \mathcal{A}_\text{L}^{+++-} \ &= \ h(s, t, u) .\\
\end{split}
\end{align}
This will be the \emph{low-energy expansion} of our theory. The functions $f$, $g$, and $h$ are polynomials of the Mandelstams given by
\begin{align}\label{eq:fgh}
\begin{split}
    f(s, t, u) \ & = \ f_2 (s^2 + t^2 + u^2) + f_3\, stu + f_4 (s^2 + t^2 + u^2)^2 + \ldots, \\
    g(s| t, u) \ & = \ g_2 s^2 + g_3 s^3 + g_{4,1} s^4 + g_{4,2} s^2(s^2 + t^2 + u^2) + \ldots, \\
    h(s, t, u) \ & = \ h_3\, stu + h_5\, stu (s^2 + t^2 + u^2) + \ldots.
\end{split}
\end{align}
These functions are fixed by the symmetries. $f$ includes every term which is totally symmetric in $s$, $t$, and $u$. $h$ is also symmetric, and includes every monomial which is $s t u$ times a symmetric term. $g(s|t, u)$ is $s^2$ times every monomial which is symmetric in $t$ and $u$. They represent non-renormalizable interactions characteristic of the EFT picture we use here. The Lagrangian that leads to a particular amplitude is not unique in general because field redefinitions may change the Lagrangian but may not change the amplitudes. As an example, the amplitudes \eqref{eq:fgh} can be found from
\begin{align}\label{eq:Lagrwitha1a2}
    \mathcal{L} &= -\frac{1}{4} F_{\mu \nu} F^{\mu \nu} +\frac{g_2 + f_2}{16} \big(F_{\mu \nu}F^{\mu \nu}\big)^2 +\frac{g_2 - f_2}{16}  \big(F_{\mu \nu}\tilde F^{\mu \nu}\big)^2 +\ldots \, ,
\end{align}
quoted in the introduction. In general, the constants $f_k$, $g_{k(,i)}$ and $h_k$ will multiply operators containing $2k$ derivatives, and we will therefore refer to these constants as $2k$-derivative coefficients.

\subsection{Partial wave expansion}

Next in this analysis is to understand which of these helicity amplitudes can be used to derive constraints. This hinges on them having a partial wave expansion with certain positivity properties. We may, of course, take linear combinations of amplitudes -- in fact, we will need to do so to ensure that our observables retain the $s\leftrightarrow t$ crossing symmetry that is necessary for the contour deformation argument in the next section. 

The general method of analyzing unitarity constraints in the context of spinning external operators was nicely outlined in \cite{Hebbar:2020ukp} (see also \cite{Bellazzini2016talk, deRham:2017zjm} for previous discussion on constraints from unitarity and crossing for external particles with spin). The partial wave expansion for spinning external particles with helicities $\lambda_i$ takes the form \cite{Jacob:1959at}
\begin{align}
    \mathcal{A}^{\lambda_1 \lambda_2 \lambda_3 \lambda_4} = \sum_J 16 \pi (2 J + 1) A^i_J(s) d^J_{\lambda_1 - \lambda_2, \lambda_4 - \lambda_3}(\theta)  \, ,
\end{align}
where $\cos\theta=1+\frac{2u}s$, and where the Wigner d-matrices $d_{a, b}$ generalize the Legendre polynomials to spinning external states. The index $i$ depends on the external helicities. Explicitly, the expansion for each of our five amplitudes reads:
\begin{align}
    \begin{split}
    \mathcal{A}_\text{H}^{++++} \ &= \ \sum_{J = 0, 2, 4, \ldots} 16 \pi   (2 J + 1) {A}^1_J(s) d^J_{0,0}(\theta)\,, \\
    \mathcal{A}_\text{H}^{++--} \ &= \  \sum_{J = 0, 2, 4, \ldots} 16 \pi  (2 J + 1) {A}^2_J(s) d^J_{0,0}(\theta)\,, \\
    \mathcal{A}_\text{H}^{+--+} \ &= \  \sum_{J =  2, 3, 4, \ldots} 16 \pi  (2 J + 1) {A}^3_J(s) d^J_{2,2}(\theta)\,, \\
    \mathcal{A}_\text{H}^{+-+-} \ &= \  \sum_{J =  2, 3, 4, \ldots} 16 \pi  (2 J + 1) {A}^4_J(s) d^J_{2,-2}(\theta)\,, \\
    \mathcal{A}_\text{H}^{+++-} \ &= \   \sum_{J =  2, 4, 6, \ldots} 16 \pi  (2 J + 1) {A}^5_J(s) d^J_{0,2}(\theta)\,.
\end{split}
\end{align}

Unitarity leads to some positivity properties on the imaginary part of these partial waves. These are called the \emph{spectral densities}, defined by $\rho^i_J = \IM A^i_J$. Unitarity implies (see appendix \ref{app:unitarity} for details): 
\begin{align}
    \rho^2_J \pm \rho^1_J \ \geqslant \ 0, \qquad J = 0, 2, 4, \ldots ,\\
    \rho^3_J \ \geqslant \ 0, \qquad J = 2, 3, 4, \ldots,
\end{align}
plus the condition
\begin{align}
    \rho^4_J = (-1)^J \rho^3_J \, .
\end{align}
As a result, we have four positive combinations of spectral densities: 
\begin{align}
    \rho^2_J \pm \rho^1_J \ \geqslant \ 0, \qquad J = 0, 2, 4, \ldots ,\\
    \rho^3_J + \rho^4_J \ \geqslant \ 0, \qquad J = 2, 4, 6, \ldots, \\
    \rho^3_J - \rho^4_J\ \geqslant \ 0, \qquad J = 3, 5, 7, \ldots,
\end{align}
It is interesting that $\rho_5$ does not participate in any of the positivity requirements derived in appendix \ref{app:unitarity}. This ultimately arises from the fact that $A_J^5$ appears in these constraints quadratically. The resulting constraint, $0\leqslant |A_J^5|^2 \leqslant 1/4$, is trivially satisfied. Constraints may be imposed on $\rho_5$ by considering a more general unitarity setup such as that of \cite{Bern:2021ppb}, or by abandoning the weak-coupling assumption.

\section{Sum Rules from Dispersion Relations}
\label{sec:three}

We have now established the low- and high-energy parameterizations of our amplitudes using the EFT and the partial wave expansions, respectively. 
Dispersion relations relate these by requiring that a contour integral over the entire amplitude in the complex $s$-plane vanishes. This requires the following assumptions:
\begin{enumerate}
    \item \emph{Causality}: for fixed $u<0$,  $\mathcal{A}(s, u)$ is analytic on the upper half-plane, $\text{Im}(s) >0$.
    \item \emph{Regge boundedness}: for fixed $u<0$, the total amplitude falls off faster than $s^2$ for large $s$. Specifically, 
    \begin{equation}
    \label{eq:Reggebound}
    \lim_{|s| \to \infty} \mathcal{A}(s, u) / s^2 = 0.
\end{equation}
 This large-$s$ behavior has been established rigorously for scattering in gapped theories \cite{Froissart:1961ux, Martin:1965jj}, however in our case we will take it as an unproven assumption.
	\item \emph{Weak coupling}: we assume that the low-energy amplitudes are weakly coupled at least up to the scale of new physics $M$, so that low-energy loops are suppressed. As a result, $M$ is the lowest energy where cuts could possibly appear in the amplitude. In general, we will get stronger bounds with a higher value of $M$, and weaker bounds with a lower, or ``more conservative'' value of $M$.
\end{enumerate}

We define the amplitude on the lower half-plane by analytic continuation, \emph{i.e.}\ $\mathcal{A}(s^*, u^*) := \mathcal{A}^*(s, u)$. So the first assumption implies that the amplitude is analytic everywhere but the real $s$-line.

These assumptions imply what we call a \emph{doubly-subtracted dispersion relation}:
\begin{figure}
     \centering
         \includegraphics{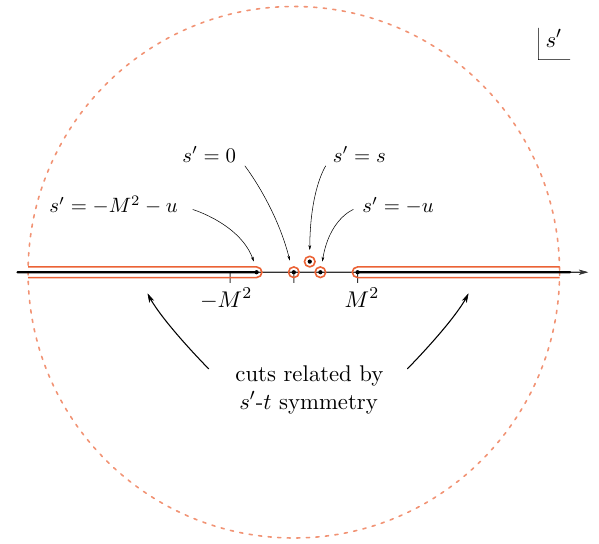}
        \caption{The dispersion relation used in \eqref{eq:doubledisp} starts from the contour at infinity (dashed). The contour is deformed inwards, picking up contributions from the three poles indicated and from two cuts on the real $s'$ axis, starting at $M^2$ and $-M^2-u$ respectively. Based on the assumed symmetry $s'\leftrightarrow t=-s'-u$, the integrals over the two cuts can be combined.}
         \label{fig:contour}
\end{figure}
\begin{align}
\label{eq:doubledisp}
    \oint_\infty \frac{ds'}{2 \pi i (s - s')} \frac{\mathcal{A}(s',u)}{s'(s' + u)} =0 .
\end{align}
Doubly-subtracted refers to the two additional powers of $s'$ in the denominator compared to the factor $(s-s')^{-1}$.
Now we assume that, at low energies, the amplitude is given by our EFT expansion, and at high energies, it is given by the partial wave expansion. The integral in~\eqref{eq:doubledisp} obtains contributions from the poles at $s' = 0$, $s' = s$, and $s' = -u$, and from cuts on the real-$s$ axis, starting at $s' = M^2$ and $s' = -M^2 - u$. In this paper, we will consider only $s-t$ symmetric amplitudes, which allows us to relate the $s' = s$ pole to the $s' = -u$ pole, and the left-handed cut to the right-handed cut (similar to the set-up in \cite{Caron-Huot:2021rmr}). We visualize the analytic structure and the combination of cuts in figure~\ref{fig:contour}. As a result, the dispersion relation becomes
\begin{align}\label{eq:disprel}\nonumber
       & \frac{\mathcal{A}_\text{L}(s, u)}{s (s + u)} + \underset{s' = 0}{\text{Res} } \left[ \frac{(2 s' + u) \mathcal{A}_\text{L}(s', u)}{s'(s' + u)(s'-s)(s' + s + u)} \right] 
        \\&\qquad\qquad\qquad = \int_{M^2}^\infty \frac{ds'}{\pi} \text{Im} \left[  \frac{(2 s' + u) \mathcal{A}_\text{H}(s', u)}{s'(s' + u)(s'-s)(s' + s + u)} \right].
\end{align}

\subsection{Sum rules}
\label{sec:sumrules}
The fact that the contour integral isolates the imaginary part of the partial wave expansion is key to this argument. 
It means that we can derive positivity bounds on the low-energy amplitudes, and thus the EFT coefficients, by choosing observables which have positive partial wave expansions.

We will consider general $s\leftrightarrow t$ symmetric linear combinations of the amplitudes defined by 
\begin{equation}\label{eq:Ax1x2}
    \mathcal A[x_1,x_2]=x_1\, \mathcal A^{++++}+\mathcal A^{++--}+\mathcal A^{+--+} + x_2\, \mathcal A^{+-+-},
\end{equation}
which has a low-energy expansion
\begin{equation}
    \mathcal A_{\mathrm L}[x_1,x_2](s,t,u)=x_1\,f(s,t,u)+g(s|t,u)+g(t|s,u)+x_2\,g(u|s,t).
\end{equation}

The unitarity considerations in section \ref{sec:setup} imply that $\mathcal A_{\mathrm H}[x_1,x_2]$ satisfies appropriate positivity conditions for $x_1\in[-1,1]$ and any $x_2$. Let us see what sum rules result from considering these amplitudes. 

Starting from \eqref{eq:Ax1x2} and using the dispersion relation \eqref{eq:disprel} we get
\begin{align}
\begin{split}
    & 2 \,  g_{2} + 2 \,  x_1 \, f_2 - 3 \, u \, g_{3} - u \,  x_1 \, f_3 + \ldots \\
    &  = \left\langle N \, d^J_{0,0}(\theta) \right\rangle_{x_1}  + \left\langle N \left[ d^J_{2,2}(\theta) + x_2 d^J_{2,-2}(\theta)  \right] \right\rangle_{e} + \left\langle N \left[ d^J_{2,2}(\theta) - x_2 d^J_{2,-2}(\theta)  \right] \right\rangle_{o},
    \label{eq:master_sum_rule}
\end{split}
\end{align}
where 
\begin{align}
    N = \frac{(2 m^2 + u)}{(m^2 + u)(m^2 - s)(m^2 + s + u)},
\end{align}
and where we have defined the brackets
\begin{align}
\begin{split}
\left\langle\cdots\right\rangle_{x_1}&=\sum_{J=0,2,\ldots}16\pi(2J+1)\int_{M^2}^\infty \frac{dm^2}{m^2}(\rho^2_J+x_1\rho^1_J)(\ldots),
\\
\left\langle\cdots\right\rangle_{e}&=\sum_{J=2,4,\ldots}16\pi(2J+1)\int_{M^2}^\infty \frac{dm^2}{m^2}\rho^3_J(\ldots),
\\
\left\langle\cdots\right\rangle_{o}&=\sum_{J=3,5,\ldots}16\pi(2J+1)\int_{M^2}^\infty \frac{dm^2}{m^2}\rho^3_J(\ldots).
\label{eq:bracketsdefinition}
\end{split}
\end{align}
For $x_1\in[-1,1]$, all of these brackets represent an integral over a positive measure for all $J$ in the respective sums. In the following, it will be convenient to define
\begin{align}
\begin{split}
       \langle\cdots\rangle_+  & =  \langle\cdots\rangle_{x_1}\big|_{x_1 = 1}, \\
   \langle\cdots\rangle_0  & =  \langle\cdots\rangle_{x_1}\big|_{x_1 = 0},\\
   \langle\cdots\rangle_-  & =  \langle\cdots\rangle_{x_1}\big|_{x_1 = -1}.
   \label{eq:bracketplusmindefinition}
\end{split}
\end{align}

Now we are in a position to derive multiple sum rules from the ``master sum rule'' (\ref{eq:master_sum_rule}). We do this with a double expansion in the small $s$ and small $u$ limit. Due to the symmetry $s \leftrightarrow -s - u$, we follow \cite{Caron-Huot:2021rmr} and replace the expansion in $s$ with the more convenient expansion in $s(s + u)$. At each order in $s(s+u)$ we get a sum rule $\mathcal C_{2n,u}[x_1,x_2]$, making \eqref{eq:master_sum_rule} equivalent to
\begin{align}
    \sum_{n = 1} [s(s + u)]^{n-1} \mathcal{C}_{2n, u}[x_1, x_2] = 0.
\end{align}

\subsection{Four-, six-, and eight-derivative operators}

To isolate the dependence on the low-energy EFT coefficients, we next expand the sum rules $\mathcal{C}_{2n, u}[x_1, x_2]$ at small $u$, or equivalently take $u$-derivatives followed by setting $u = 0$. Let us look at the results of doing this for the first few orders in the $s$ and $u$ expansion:
\begin{align}\label{eq:sumrule20}
\begin{split}
    \mathcal{C}_{2, 0}  &=  2 x_1 f_2 + 2 g_{2}  - \left\langle \frac{2}{m^4} \right\rangle_{\!x_1} - \left\langle \frac{2}{m^4} \right\rangle_{\!e} - \left\langle \frac{2}{m^4} \right\rangle_{\!o} \, , \\ \,
    \mathcal{C}_{2, 0}'  &=  - x_1 f_3 - 3 g_{3} - \left\langle \frac{2 \mathcal{J}^2-3}{m^6} \right\rangle_{\!x_1} - \left\langle \frac{2 \mathcal{J}^2-11}{m^6} \right\rangle_{\!e} - \left\langle \frac{2 \mathcal{J}^2-11}{m^6} \right\rangle_{\!o} \,,\\
    \mathcal{C}_{2, 0}''  &=  8 x_1 f_4 + 4 g_{4,1} + (6 + 2 x_2) g_{4,2}  - \left\langle \frac{\mathcal{J}^4-8 \mathcal{J}^2+8}{2 m^8} \right\rangle_{\!x_1}\nonumber \\
    &\quad  - \left\langle \frac{(6 + x_2) \mathcal J^4  - (96 + 2x_2)\mathcal J^2 + 360}{12m^8} \right\rangle_{\!e} - \left\langle (x_2 \to - x_2)  \right\rangle_{o}\,,\\
    \mathcal{C}_{4, 0}  &=  4 x_1 f_4 + 2 g_{4,1} + 4 g_{4,2} - \left\langle \frac{2}{m^8} \right\rangle_{\!x_1} - \left\langle \frac{2}{m^8} \right\rangle_{\!e} - \left\langle \frac{2}{m^8} \right\rangle_{\!o}\,,
\end{split}
\end{align}
where $\mathcal J^2=J(J+1)$.
Using these amplitudes, we can solve for the EFT coefficients in terms of the brackets. For example, 
\begin{align}
    f_2 \ &= \ \left\langle \frac{1}{2m^4} \right\rangle_{\!+} - \left\langle \frac{1}{2m^4} \right\rangle_{\!-} \, , \\
    g_2 \ &= \ \left\langle \frac{1}{2m^4} \right\rangle_{\!+} + \left\langle \frac{1}{2m^4} \right\rangle_{\!-}  + \left\langle \frac{1}{m^4} \right\rangle_{\!e}  +  \left\langle \frac{1}{m^4} \right\rangle_{\!o} \, . 
\end{align}
In table \ref{tab:EFTbrackets} in appendix~\ref{app:appendixsumrules}, we collect the sum rules for all the seven EFT coefficients that are analyzed in this paper: $f_2$, $g_2$, $f_3$, $g_3$, $f_4$, $g_{4,1}$ and $g_{4,2}$.  

\subsection{Null constraints}

The sum rules derived above contain redundancies, the first of which happens at eight-derivative order. This allows us to derive \emph{null constraints} by writing the same coefficient two different ways \cite{Caron-Huot:2021rmr}. For example, 
\begin{align}
    f_4 \ &= \  \left\langle \frac{\mathcal{J}^4-8 \mathcal{J}^2+8}{32 m^8} \right\rangle_{+} - \left\langle \frac{\mathcal{J}^4-8 \mathcal{J}^2+8}{32 m^8} \right\rangle_{-}\,,\\
    f_4 \ &= \   \left\langle \frac{2}{8m^8} \right\rangle_{+} - \left\langle \frac{2}{8m^8} \right\rangle_{-} \, ,
\end{align}
which directly leads to the null constraint
\begin{align}
     \mathcal{X}_{2,0} = \left\langle \frac{\mathcal{J}^2 (\mathcal{J}^2-8)}{2 m^8} \right\rangle_{+} - \left\langle \frac{\mathcal{J}^2 (\mathcal{J}^2-8)}{2 m^8} \right\rangle_{-}  = 0\,.
     \label{eq:fnc1}
\end{align}
As we have removed $f_4$ from this equation, $\mathcal{X}_{2,0}$ represents a constraint on high-energy data only. Enforcing this constraint allows for the derivation of stronger numerical bounds -- essentially, instead of searching for functionals which a positive for all possible high-energy data, we can look for functionals which are positive for all possible high-energy data satisfying~\eqref{eq:fnc1}.

We can do the same for the combination $g_{4,1} + 2 g_{4,2}$, giving 
\begin{equation}
    \mathcal{Y}_{2,0} = \left\langle\frac{\mathcal{J}^2 (\mathcal{J}^2-8)}{ m^8}\right\rangle_0+
    \left\langle \frac{7 \mathcal{J}^4-98 \mathcal{J}^2+312}{6 m^8}\right\rangle_e
    +\left\langle\frac{5 \mathcal{J}^4-94 \mathcal{J}^2+312}{6 m^8}\right\rangle_o = 0
    \label{eq:gnc1}.
\end{equation}
These null constraints ultimately derive from crossing symmetry, which restricts the number of possible EFT coefficients.\footnote{In fact, crossing symmetry of the $h(s, t, u)$ function also implies null constraints, but since $\rho_J^5$ is not manifestly positive, these are not useful.} The result is that there are more sum rules than coefficients, hence the redundancy we see above. 
The null constraints inherently apply to only the high-energy part of the amplitude; in essence they bound the partial waves, and seem to imply a weak form of low spin dominance, similar to what is considered in \cite{Bern:2021ppb}. 

\subsubsection{Systematics for null constraints}

Though we will only analyze the seven operators with eight or fewer derivatives, we will need to consider higher sum rules in order to generate more null constraints. In practice, it is possible to generate hundreds of null constraints from these formulas in Mathematica. 

Such null constraints are most straightforwardly computed by considering completely $s$-$t$-$u$ symmetric amplitudes. We have two linearly independent amplitudes satisfying this property: 
\begin{align}\label{eq:amplfornull1}
    \frac{1}{2}(\mathcal{A}[1, 1] - \mathcal{A}[-1, 1]) = \mathcal{A}^{++++} \, ,
\end{align}
and 
\begin{equation}\label{eq:amplfornull2}
\mathcal A[0,1]=\mathcal A^{++--}+\mathcal A^{+--+}+\mathcal A^{+-+-} \, .
\end{equation}

Crossing symmetry simplifies the low-energy expansions to 
\begin{align}
\begin{split}
    \mathcal A^{++++}_{\mathrm L}(s,t,u) &=
    f_2(s^2+t^2+u^2)+f_3 stu + f_4 (s^2+t^2+u^2)^2+\ldots \, , \\
    \mathcal A_{\mathrm L}[0,1](s,t,u) &=
    g_2(s^2+t^2+u^2)+3g_3stu+\frac{g_{4,1}+2g_{4,2}}2(s^2+t^2+u^2)^2+\ldots.
\end{split}
\end{align}

We shall refer to null constraints that arise from the first amplitude \eqref{eq:amplfornull1} as ``$f$-type'', and those from the second amplitude \eqref{eq:amplfornull2} as ``$g$-type''. These both arise from completely $s$-$t$-$u$ symmetric polynomials.
It is easy to realize that the most general such amplitude can be written as a linear combination of simple polynomials $(s^2+t^2+u^2)^a(stu)^b$. 
The structure of the low-energy amplitude is therefore completely equivalent to the one considered in \cite{Caron-Huot:2020cmc,Caron-Huot:2021rmr}.

The high-energy expansion is different. In \cite{Caron-Huot:2021rmr}, general null constraints were written down on the form
\begin{align}\nonumber
0&=
\mathcal X^{\mathrm{scalar}}_{k,u}
=\Bigg\langle \frac{2m^2+u}{um^2(m^2+u)}\frac{m^2\mathcal P_J(1+\frac{2u}{m^2})}{(um^2(m^2+u))^{k/2}}
\\
&\quad -\res_{u'=0}\frac{(2m^2+u')(m^2-u')(m^2+2u')}{m^2(u-u')u'(m^2-u)(m^2+u')(m^2+u+u')}\frac{m^2\mathcal P_J(1+\frac{2u'}{m^2})}{(u'm^2(m^2+u'))^{k/2}}\Bigg\rangle_{\mathrm{scalar}}\,,
\end{align}
where $\mathcal P_J(\cos\theta)$ are Gegenbauer polynomials (which reduce to Legendre polynomials in four dimensions).
Since we have exactly the same cancellations on the low-energy side as in that paper, we can find $f$-type null constraints in the case of photon scattering by replacing
\begin{align}
\left\langle \cdots \mathcal P_J(\cos\theta)\right\rangle_{\mathrm{scalar}}&\longrightarrow \left\langle\cdots d_{0,0}^{J}(\theta)\right\rangle_+ - \left\langle\cdots d_{0,0}^{J}(\theta)\right\rangle_- \,. 
\label{eq:modifynullconstraintsf}
\end{align}
Likewise, the $g$-type null constraints follow from
\begin{align}
\begin{split}
    \left\langle \cdots \mathcal P_J(\cos\theta)\right\rangle_{\mathrm{scalar}}&\longrightarrow \left\langle\cdots d_{0,0}^{J}(\theta)\right\rangle_0 
\\&\qquad+\left\langle\cdots [d_{2,2}^{J}(\theta)+d_{2,-2}^{J}(\theta)]\right\rangle_e+\left\langle\cdots [d_{2,2}^{J}(\theta)-d_{2,-2}^{J}(\theta)]\right\rangle_o.
\end{split}
\label{eq:modifynullconstraintsg}
\end{align}
Note that the $\mathcal P_J(\cos\theta)=d^J_{0,0}(\theta)$, so that the expressions entering in the $0$ brackets are identical to those entering in the scalar bracket of \cite{Caron-Huot:2020cmc,Caron-Huot:2021rmr}.

\section{Results}
\label{sec:results}

In this section, we report what comes out of the dispersion relation methods above. 
We are interested in analyzing all coefficients for operators with up to eight derivatives. We will display a variety of analytic and numerical bounds. 

\subsection{Bounds on four-derivative coefficients}
\label{subsec:Bounds4deriv}

Recall from section~\ref{sec:three} that 
\begin{align}
\begin{split}
    f_2 \ &= \ \left\langle \frac{1}{2m^4} \right\rangle_{\!+} - \left\langle \frac{1}{2m^4} \right\rangle_{\!-} \, , \\
    g_2 \ &= \ \left\langle \frac{1}{2m^4} \right\rangle_{\!+} + \left\langle \frac{1}{2m^4} \right\rangle_{\!-}  + \left\langle \frac{1}{m^4} \right\rangle_{\!e}  +  \left\langle \frac{1}{m^4} \right\rangle_{\!o} \, . 
    \end{split}
\end{align}
As a result, we find that $g_2 + f_2 = \langle m^{-4} \rangle_++\langle m^{-4} \rangle_e+\langle m^{-4} \rangle_o$ and $g_2 - f_2 = \langle m^{-4}\rangle_-+\langle m^{-4} \rangle_e+\langle m^{-4} \rangle_o$. Since all the involved brackets, defined by \eqref{eq:bracketsdefinition} and \eqref{eq:bracketplusmindefinition}, are manifestly non-negative, we are left with two positivity conditions, 
\begin{align}
\begin{split}
    g_2+f_2&\geqslant0,
    \\
    g_2-f_2&\geqslant0,
    \label{eq:positivityg2f2}
\end{split}
\end{align}
or equivalently $|f_2|\leqslant g_2$. This shows that the two constants $a_1$ and $a_2$ defined in \eqref{eq:Lagrwitha1a2intro} are individually non-negative: $a_1\geqslant0$, $a_2\geqslant0$. In \cite{Cheung:2014ega}, this conclusion was reached by assuming an ansatz for the UV completion and requiring the absence of tachyons and ghosts.  The more general dispersion relation method employed in that paper only implied the result $a_1+a_2\geqslant0$, equivalent to $g_2\geqslant0$. It was later observed using dispersion relations in \cite{Bellazzini2016talk, Falkowski}.

\subsection{Null constraints and bounds on six-derivative coefficients}

Next we move to coefficients $g_3$ and $f_3$ appearing at six-derivative order. These are given by
\begin{align}
\begin{split}
    f_3 \ &= \ \left\langle \frac{3- 2 \mathcal{J}^2}{2m^6} \right\rangle_{\!+} - \left\langle \frac{3- 2 \mathcal{J}^2}{2m^6} \right\rangle_{\!-} \, , \\
    g_3 \ &= \ \left\langle \frac{3- 2 \mathcal{J}^2}{6m^6} \right\rangle_{\!+} + \left\langle \frac{3- 2 \mathcal{J}^2}{6m^6} \right\rangle_{\!-}  + \left\langle \frac{11- 2 \mathcal{J}^2}{3m^6} \right\rangle_{\!e}  +  \left\langle \frac{11- 2 \mathcal{J}^2}{3m^6} \right\rangle_{\!o} \, . 
    \label{eq:f3g3main}
    \end{split}
\end{align}
where $\mathcal J^2=J(J+1)$.

The considerations in section~\ref{sec:three} will allow us to obtain analytically a two-sided bound on the ratio $g_3/g_2$.\footnote{
From this point and onwards, we will assume that $g_2>0$.
} We then consider numerically the problem of bounding all coefficients $g_2$, $f_2$, $g_3$, $f_3$.

\subsubsection[Two-sided analytic bounds]{Two-sided analytic bounds}

We can derive two-sided bounds analytically using the bracket definitions of $g_2$ and $g_3$, plus the first $g$-type null constraint, $\mathcal{Y}_{2,0}$:
\begin{align}
\label{eq:normalizationrulemain}
    g_2&=\left\langle\frac1{m^4}\right\rangle_{\!0}+\left\langle
    \frac1{m^4}
    \right\rangle_{\!e}+\left\langle\frac1{m^4}\right\rangle_{\!o} \, ,
    \\
    \label{eq:g3sumrulemain}
    g_3&=\left\langle 
    \frac{3 - 2\mathcal J^2}{3m^6}
    \right\rangle_{\!0}+\left\langle  
     \frac{11-2\mathcal J^2}{3m^6}
    \right\rangle_{\!e}+\left\langle 
     \frac{11-2\mathcal J^2}{3m^6}
    \right\rangle_{\!o} \, ,
   \\
   \label{eq:nullWmain}
   0& =\left\langle
   \frac{\mathcal J^2(\mathcal J^2-8)}{ m^8}
   \right\rangle_{\!0}+\left\langle 
    \frac{7 \mathcal{J}^4-98 \mathcal{J}^2+312}{6 m^8}
   \right\rangle_{\!e}+\left\langle 
   \frac{5 \mathcal{J}^4-94 \mathcal{J}^2+312}{6 m^8}
   \right\rangle_{\!o} \, ,
\end{align}
We will show that these sum rules lead to a two-sided bound on $g_3/g_2$,
\begin{equation}\label{eq:boundg3g2}
  \frac{r_{\mathrm{min}}}{M^2}
\leqslant \frac{g_3}{g_2}\leqslant \frac{1}{M^2},\qquad 
r_{\mathrm{min}} = -\left(\frac{13}6+\sqrt{\frac{7877}{996}}\right)
 =-4.978896 .
\end{equation}

\paragraph{Upper bound:}Consider the terms entering the brackets in \eqref{eq:g3sumrulemain}. First, note that $\mathcal{J} = J(J+1)$, so
\begin{align}
\begin{split}
    \frac{3 - 2 \mathcal{J} }{3m^6}  &\leqslant
    \frac{1 }{m^6} \qquad \qquad J = 0, 2, 4,\ldots, \\
    \frac{11 - 2 \mathcal{J} }{3m^6} & \leqslant
    \frac{1 }{m^6}  \qquad \qquad  J = 2,3, 4, \ldots,
\end{split}
\end{align}
Second, note that for any $\rho_J(m^2)\geqslant0$, we have that
\begin{align}
    \int^\infty_{M^2} \frac{dm^2}{m^2}\rho_J(m^2)\frac{1}{m^6} \leqslant  \frac{1}{M^2} \int^\infty_{M^2} \frac{dm^2}{m^2}\rho_J(m^2)\frac{1}{m^4} \,, 
    \end{align}
    which implies that
    \begin{align}
    \left\langle 
    \frac{1 }{m^6} \right\rangle \leqslant   \frac{1}{M^2} \left\langle 
    \frac{1 }{m^4} \right\rangle 
\,.\end{align}
Combining these results we find that $g_3$ is bound by $g_2$:
\begin{equation}
\label{eq:upperboundg3}
    g_3\leqslant\frac1{M^2}\left(\left\langle\frac1{m^4}\right\rangle_{\!0}+\left\langle\frac1{m^4}\right\rangle_{\!e}+\left\langle\frac1{m^4}\right\rangle_{\!o}
  \,  \right) = \frac{g_2}{M^2},
\end{equation}
showing the upper bound in \eqref{eq:boundg3g2}.

\paragraph{Lower bound:} Now we must include the null constraint \eqref{eq:nullWmain}. The problem has a geometric interpretation. We will bound the ratio $g_3/g_2$ and we can therefore think of the sum rule for $g_2$, \eqref{eq:normalizationrulemain}, as a normalization. The right-hand sides of the remaining two sum rules \eqref{eq:g3sumrulemain} and \eqref{eq:nullWmain} then describe a convex hull of points in a two-dimensional space where the allowed values of $g_3/g_2$ are given by the intersection of this convex hull with the $x$-axis. The minimal value \eqref{eq:boundg3g2} is found at a point on the boundary of the convex hull, with spectral densities supported only on the odd bracket at spin $J=3$ and $J=5$. We give more details on this, as well as a graphical visualization, in appendix~\ref{app:simpleanalyicbound}.

\paragraph{$\boldsymbol{f_2}$ dependent bound:} In fact, we can immediately extend the result just described. Consider the derivation of the result \eqref{eq:boundg3g2} using an arbitrary value of $x_1$. The positivity conditions and argument are unchanged, with $\langle\cdots\rangle_0\to\langle\cdots\rangle_{x_1}$, leading to the more general result
\begin{equation}\label{eq:g3plusx1f3}
    \frac{r_{\mathrm{min}}}{M^2}\leqslant \frac{g_3+\frac{x_1}3f_3}{g_2+x_1f_2}\leqslant \frac{1}{M^2}.
\end{equation}

\subsubsection[Numerical bounds]{Numerical bounds}

We will now proceed with a numerical implementation, using optimization with the semi-definite programming solver \texttt{SDPB} \cite{Simmons-Duffin:2015qma,Landry:2019qug}. We leave a more general discussion for appendix~\ref{subsec:OptProblem}; in the present section we will briefly describe an example with $g_2$ and $g_3$.

Let us use the results from \eqref{eq:normalizationrulemain}--\eqref{eq:nullWmain} to define
\begin{align}
\begin{split}
\vec V^0 (m^2 ,J) &:=\left(\frac1{m^4},\,\frac{3 - 2\mathcal J^2}{3m^6},\, \frac{\mathcal{J}^2 (\mathcal{J}^2-8)}{ m^8}\right)\,,\\
\vec V^e (m^2 ,J) &:=\left(\frac1{m^4},\,\frac{11-2\mathcal J^2}{3m^6},\, \frac{7 \mathcal{J}^4-98 \mathcal{J}^2+312}{6 m^8}\right)\,, \\
\vec V^o (m^2 ,J) &:=\left(\frac1{m^4},\,\frac{11-2\mathcal J^2}{3m^6},\, \frac{5 \mathcal{J}^4-94 \mathcal{J}^2+312}{6 m^8}\right)\,,
\end{split}
\end{align}
For the numerical implementation we put $M^2=1$ and restore the $M^2$ dependence in our results using dimensional analysis.

We can now schematically write semidefinite optimization problems of this form
\begin{equation}
\begin{cases}
\text{Min} & A \\
\text{s.t.} & 0 \leqslant (A,\pm1, c)\cdot \vec V^0 \quad  \forall\, m^2\geqslant M^2,\ \forall \,J=0,2,4,\ldots,\\
& 0 \leqslant (A,\pm1, c)\cdot \vec V^e \quad \forall \,  m^2\geqslant M^2,\ \forall \,J=2,4,6,\ldots,\\
& 0 \leqslant (A,\pm1, c)\cdot \vec V^o \quad \forall \,  m^2\geqslant M^2,\ \forall \,J=3,5,7,\ldots.
\end{cases}
\label{eq:UpperBound}
\end{equation}
For practical purposes, we limit to a finite set of spins, supplemented by the condition from the limit $J\to\infty$ (see appendix~\ref{subsec:OptProblem} for more details). Using that $g_2>0$, we can use this algorithm to generate two-sided bounds on $g_3/g_2$. The lower bound follows from minimizing $A$ using the plus sign, to give $g_3/g_2\geqslant -A$. Conversely, the upper bound follows from choosing the minus sign: giving $g_3/g_2\leqslant A$ (where, of course, $A$ now is different).

\begin{figure}[ht]	
  \centering
\includegraphics[width=0.45\textwidth]{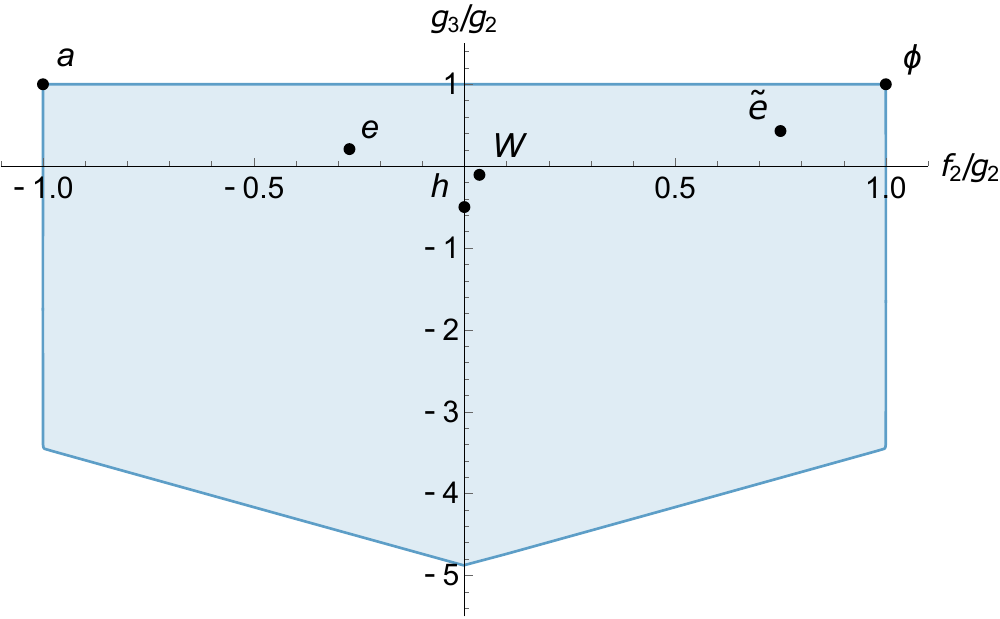}\hspace{0.1\textwidth}\includegraphics[width=0.45\textwidth]{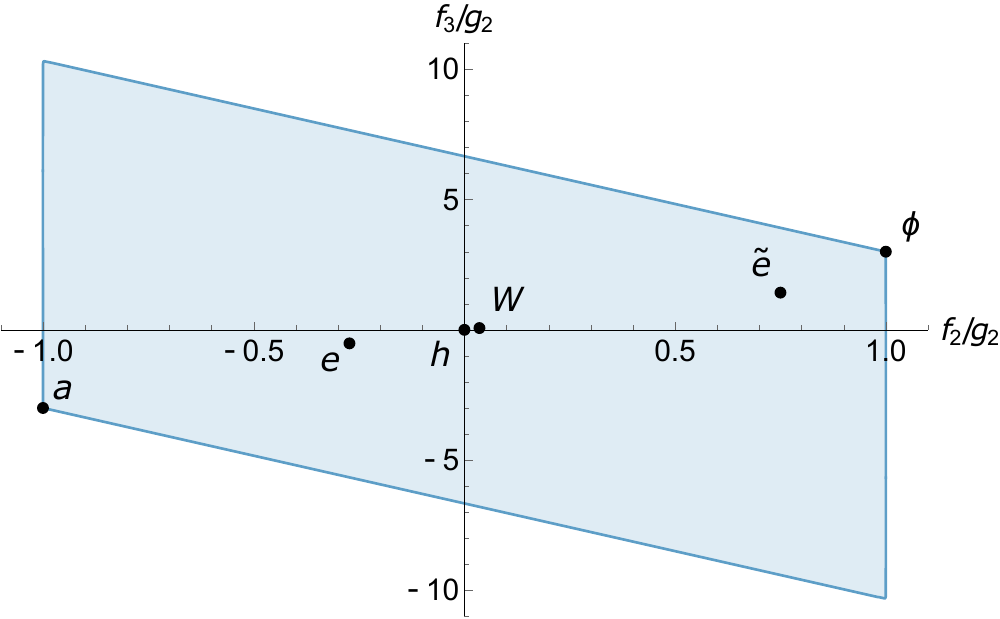}
\caption{Bounds on the six-derivative terms  $f_3/g_2$ and $g_3/g_2$ as a function of $f_2/g_2$. The dots refer to the values in the partial UV completions discussed in section~\ref{sec:partialUV}: massive axion ($a$), scalar ($\phi$)and graviton ($h$), and QED ($e$), scalar QED ($\tilde e$) and $W^\pm$ sector ($W$).}\label{fig:f2-dependence}
\end{figure}

With a suitable generalization of the optimization problem (see appendix~\ref{subsubsec:2d bounds}), we are also able to generate bounds involving more than one ratio of variables. In figure~\ref{fig:f2-dependence} we display bounds in the planes spanned by $\big(\frac{f_2}{g_2},\,\frac{g_3}{g_2}\big)$ and $\big(\frac{f_3}{g_2},\,\frac{g_3}{g_2}\big)$.

\begin{figure}[ht]	
  \centering
\includegraphics[width=0.6\textwidth]{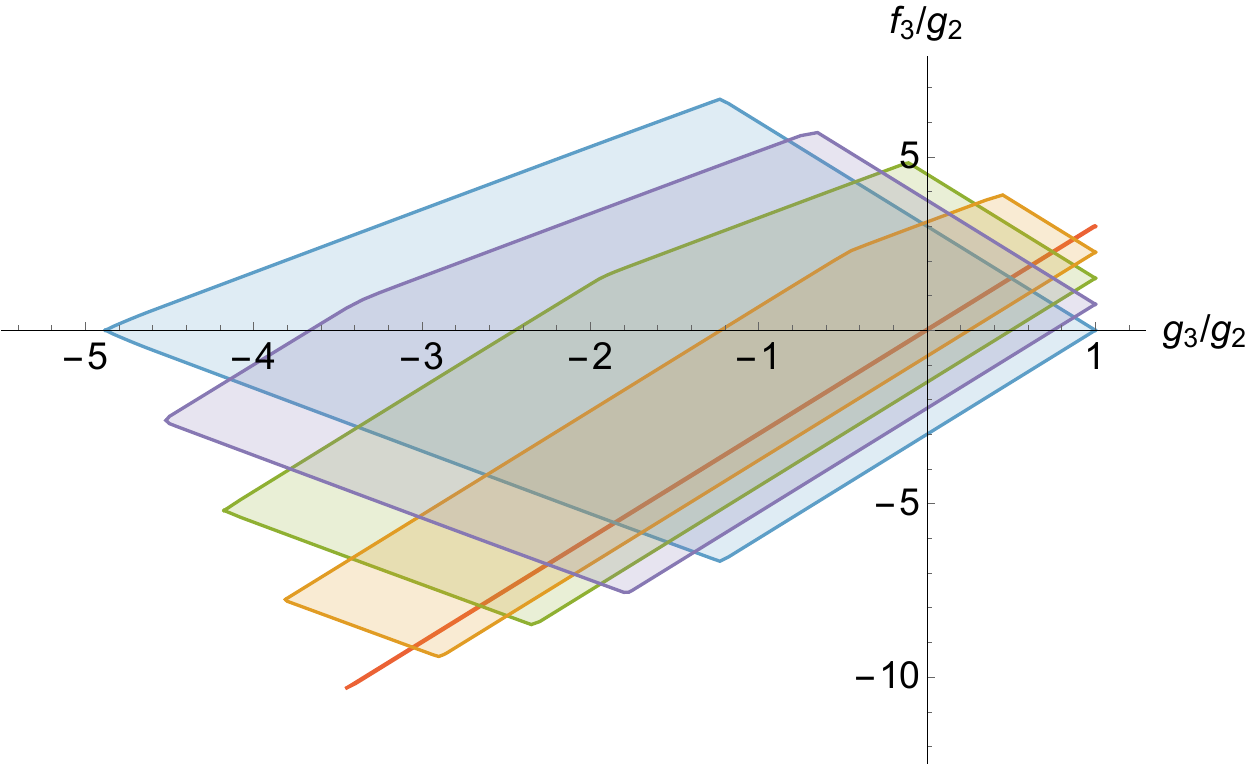}
\caption{Allowed region in the plane $(g_3/g_2,\, f_3/g_2)$ for fixed values $f_2/g_2 = k$. The  blue, purple, green, yellow and red regions correspond respectively to $ k = 0, 0.25, 0.5, 0.75, 1$. }
\label{fig:g3f3}
\end{figure}

We observe that the bounds on $g_3/g_2$ (resp.\ $f_3/g_2$) appear to be linear in $|f_2|$ (resp.\ $f_2$). A similar observation holds also for the bounds on eight-derivative coefficients considered below. 
As a consequence, the $f_2$ dependence in subsequent bounds can be captured by constructing the bound with a few different fixed values of the ratio $f_2/g_2$. In this way, the whole set of bounds involving four- and six-derivative operators can be visualized in a single two-dimensional diagram, which we give in figure~\ref{fig:g3f3}.

A special point of interest is the lower bound of $g_3$ at the special value $f_2=g_2$. We can get a simple analytic value for this point using one null constraint:
\begin{equation}
g_3\geqslant -\left(\frac32+\sqrt{\frac{2989}{720}}\right) = -3.537496, \qquad f_2=g_2.
\end{equation}
See appendix~\ref{app:f2equalsg2} for details.

\subsubsection{Dependence on the number of null constraints}

As mentioned before, two-sided bounds can be obtained by using at least one null constraint. On the other hand, we have access to an infinite family of null constraints, and it is interesting to see how our results change when including more of them.

First, note that we do not expect any improved results for upper bounds with increased number of null constraints. This can in principle be seen from the explicit form of the null constraints. However, looking ahead, it is clear that it must be the case since the upper bound is saturated by the partial UV completion given by the massive scalar and axion in table~\ref{tab:EFTcoefValues_trees} below. 
The lower bound, on the other hand, does depend on the number $f$-type null constraints, as can be seen in table~\ref{tab:Nmaxkmin}.	
\begin{table}[ht]
\centering
	\caption{Lower bounds $g_3/g_2\geqslant r_{\mathrm{min}}/M^2$ using a number $N_{g}$ ($N_f$) of $g$-type ($f$-type) null constraints.}\label{tab:Nmaxkmin}
	{
		\renewcommand{\arraystretch}{1.5}
		\begin{tabular}{|cc|c|c|}
			\hline
			$N_{g}$ &$N_{f}$ & $r_{\mathrm{min}}$& $r_{\mathrm{min}}^{f_2=g_2}$
			\\\hline
			$ 1 $& $0$ & $-4.978896$  & $-3.537496$\\
			$2 $&$0$& $-4.978896$ & $-3.537496$\\
			$ 5 $&$0$& $-4.883585$ & $-3.452668$ \\
			$ 9 $&$0$& $-4.879918 $ & $-3.449724$\\
			$ 35$ &$0$& $-4.879317 $ & $-3.448855$\\
			$ 70$ &$0$& $-4.879085 $ & $-3.448724$\\
			$145$ &$0$& $ -4.878651$ & $-3.448191$
			\\\hline
		\end{tabular}
		\quad
		\begin{tabular}{|c|c|c|}
			\hline
			$N_{g}=N_{f}$ & $r_{\mathrm{min}}$& $r_{\mathrm{min}}^{f_2=g_2}$
			\\\hline
			$ 1 $ & $-4.887208$  & $-3.537496$\\
			$2 $& $ -4.887208$ & $-3.537496$\\
			$ 5 $& $ -4.829627$ & $-3.452668$ \\
			$ 9 $ &$-4.828436$ & $-3.449724$\\
			$ 35$  &$-4.828427$ & $ -3.448855$\\
			$ 70$  &$ -4.828427$ & $-3.448724$\\
			$145$ & $-4.828427$ & $-3.448191$
			\\\hline
		\end{tabular}
	}
\end{table}
From table~\ref{tab:Nmaxkmin}, we also note that the improvement compared to a single null constraint is rather modest with respect of our needs, and for the rest of the paper we will use $N_g+N_f=9+9$ null constraints.

\subsection{Bounds on eight-derivative coefficients}

In this section we show results that involve coefficients at eight-derivative order. 
First we will consider the bounds in the $(g_{4,1}, \, g_{4,2})$ plane. 
We find that the allowed values of these couplings fall within a triangular shape, whose boundaries we determine analytically. 
We then consider bounds in the inhomogeneous space involving the coefficient $g_3$ against different combinations of eight-derivative coefficients.  

\subsubsection{Bounds involving only eight-derivative coefficients}

\begin{figure}[ht]	
  \centering
\includegraphics[width=0.8\textwidth]{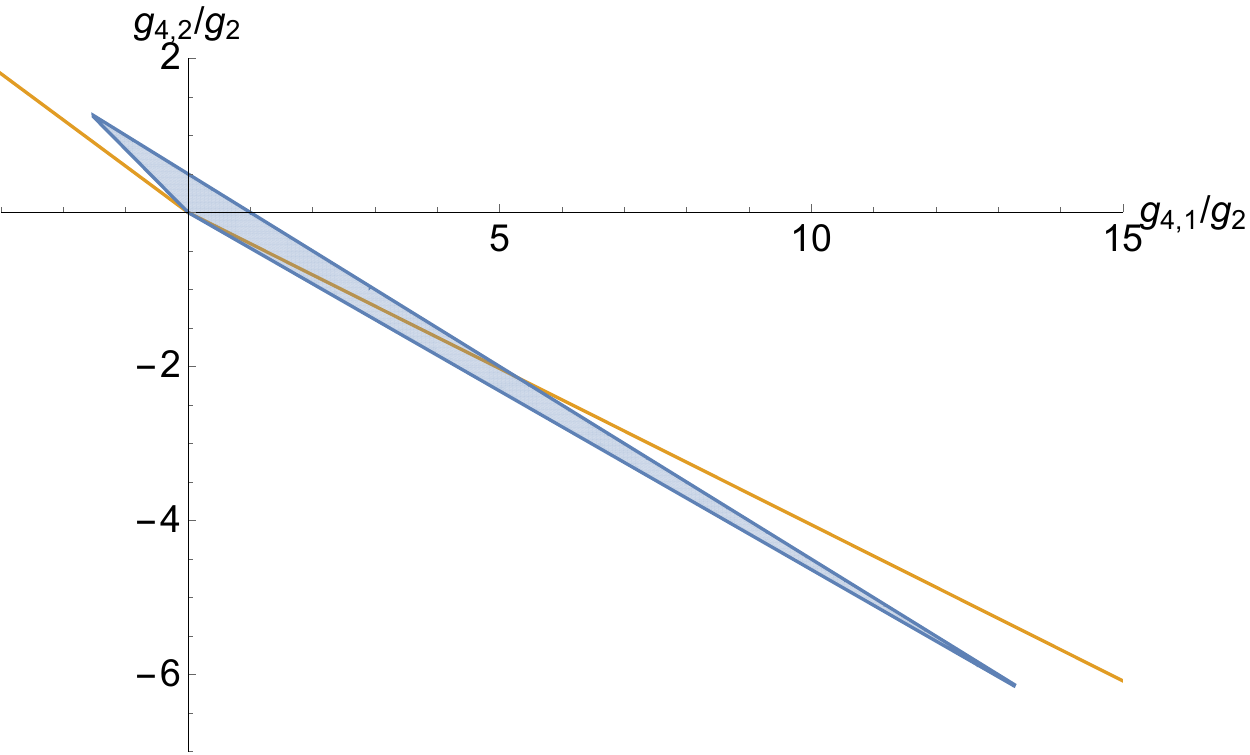}
\caption{Showing the triangle with the vertices given by \eqref{eq:triangle}, and, in orange, the bound \eqref{eq:EFTHedron} from \cite{Arkani-Hamed:2020blm}, where allowed region is above the graph.}\label{fig:g41g42}
\end{figure}

We start by considering bounds in the plane parametrized by $g_{4,1}$ and $g_{4,2}$. The relevant sum rules are
\begin{align}
g_{4,1}&=  \left\langle\frac{1}{m^8}\right\rangle_{\!0}+\left\langle\frac{-\mathcal{J}^4+2 \mathcal{J}^2+12}{12 m^8}\right\rangle_{\!e}+\left\langle\frac{\mathcal{J}^4-2 \mathcal{J}^2+12}{12 m^8}\right\rangle_{\!o} \, ,
  \\
g_{4,2}&=\left\langle\frac{\mathcal{J}^2 (\mathcal{J}^2-2)}{24 m^8}\right\rangle_{\!e}+\left\langle-\frac{\mathcal{J}^2 (\mathcal{J}^2-2)}{24 m^8}\right\rangle_{\!o}  \, .
\end{align}
An explicit analytic consideration in appendix~\ref{app:triangle} shows that the allowed values must lie in a region given by a triangle with vertices
\begin{equation}\label{eq:triangle}
(g_{4,1},g_{4,2}):\qquad (0,0),\quad \left(\frac{399}{29}\frac{g_2}{M^4},-\frac{185}{29}\frac{g_2}{M^4}\right),\quad \left(-\frac{11}{7}\frac{g_2}{M^4},\frac{9}{7}\frac{g_2}{M^4}\right),
\end{equation}
shown in figure~\ref{fig:g41g42}.

We can compare with some results from \cite{Arkani-Hamed:2020blm} for homogeneous bounds on eight-derivative coefficients. Translated to our parametrization, the bound found in \cite{Arkani-Hamed:2020blm} is
\begin{equation}\label{eq:EFTHedron}
-\frac{30}7\leqslant \frac{2g_{4,2}}{g_{4,1}+2g_{4,2}}\leqslant6,\qquad g_{4,1}+2g_{4,2}>0.
\end{equation}
Notice that our bound in \eqref{eq:triangle} is stronger than \eqref{eq:EFTHedron} in the region $g_{4,1}<0$, but weaker in the region $g_{4,1}>0$. 

We then proceed to finding a numerical bound in the $(g_{4,1},\,g_{4,2})$ plane. It turns out that the numerical bound is only slightly more constraining than the analytic bound given by the triangle \eqref{eq:triangle}, and the difference is barely visible. For instance, for $g_{4,1}/g_2>0$, even with $O(100)$ null constraints the lower bound on  $g_{4,2}/g_2$ is only stronger than the value obtained analytically from \eqref{eq:triangle} by at most $10^{-5}$.

\subsubsection{Bounds involving both six- and eight-derivative coefficients}

First, we consider the space of couplings $(g_{4,1}+2 g_{4,2})/g_2$ and $g_3/g_2$. We choose this combination of $g_{4,1}$ and $g_{4,2}$ because it corresponds to a $s$-$t$-$u$ symmetric combinations of the amplitudes. 

\begin{figure}[ht]	
  \centering
\includegraphics[width=0.8\textwidth]{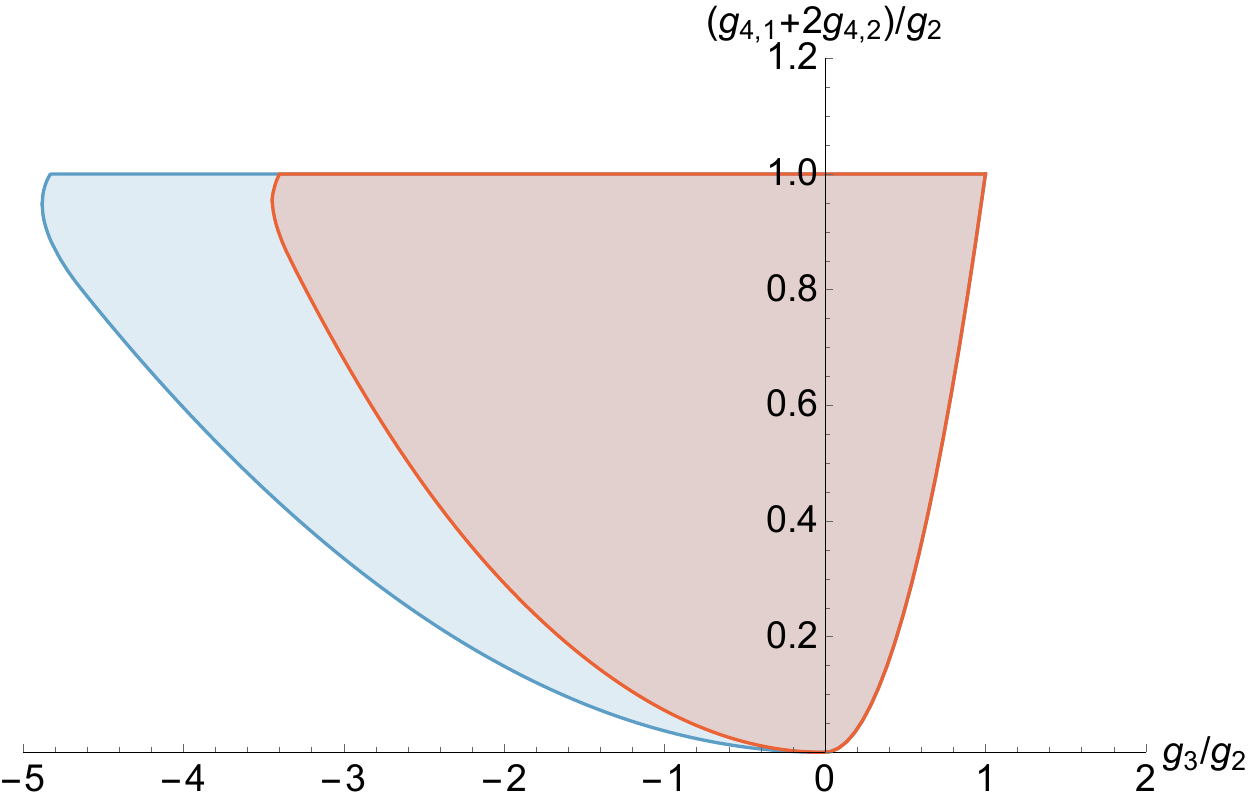}
\caption{Bounds in the plane $\frac{g_2}{M^2}e_1=g_3$, $\frac{g_2}{M^4}e_2=g_{4,1}+2g_{4,2}$ normalized to $g_2$. The blue, red regions corresponds to $f_2/g_2=0,\pm 1$. Intermediate allowed regions can be obtained by linear interpolation.}\label{fig:g3g4}
\end{figure}

Then, using the algorithm described in appendix~\ref{subsubsec:2d bounds}, we get numerically the allowed region in figure~\ref{fig:g3g4}. In particular, we consider the extra constraint $f_2/g_2=0,\pm 1$ to see how the bounds depend on this ratio.

This plot has strong similarities with the figure 8 found in \cite{Caron-Huot:2020cmc} for scalars. In fact, we believe that the amplitudes (specifically, the $s-t-u$ symmetric $\mathcal{A}[0,1]$ amplitude) for the theory saturating the bounds are the same as the amplitudes given in $(5.2)$ of that paper. We know that this is the case for the scalar / axion amplitudes, which sit at the top right, because we have computed them directly. The minimum value $g_3$ in the red region coincides (after accounting for a factor of 3 due to conventions) with the value of the corresponding region in \cite{Caron-Huot:2020cmc}. Therefore we believe that the theory that lives at the kink in the red region has an $\mathcal{A}[0,1]$ amplitude corresponding to the $stu$-pole amplitude given in $(5.2)$ in that paper. We do not know what amplitudes lie at the blue kink on the upper left.

\begin{figure}[ht]	
  \centering
\includegraphics[width=0.6\textwidth]{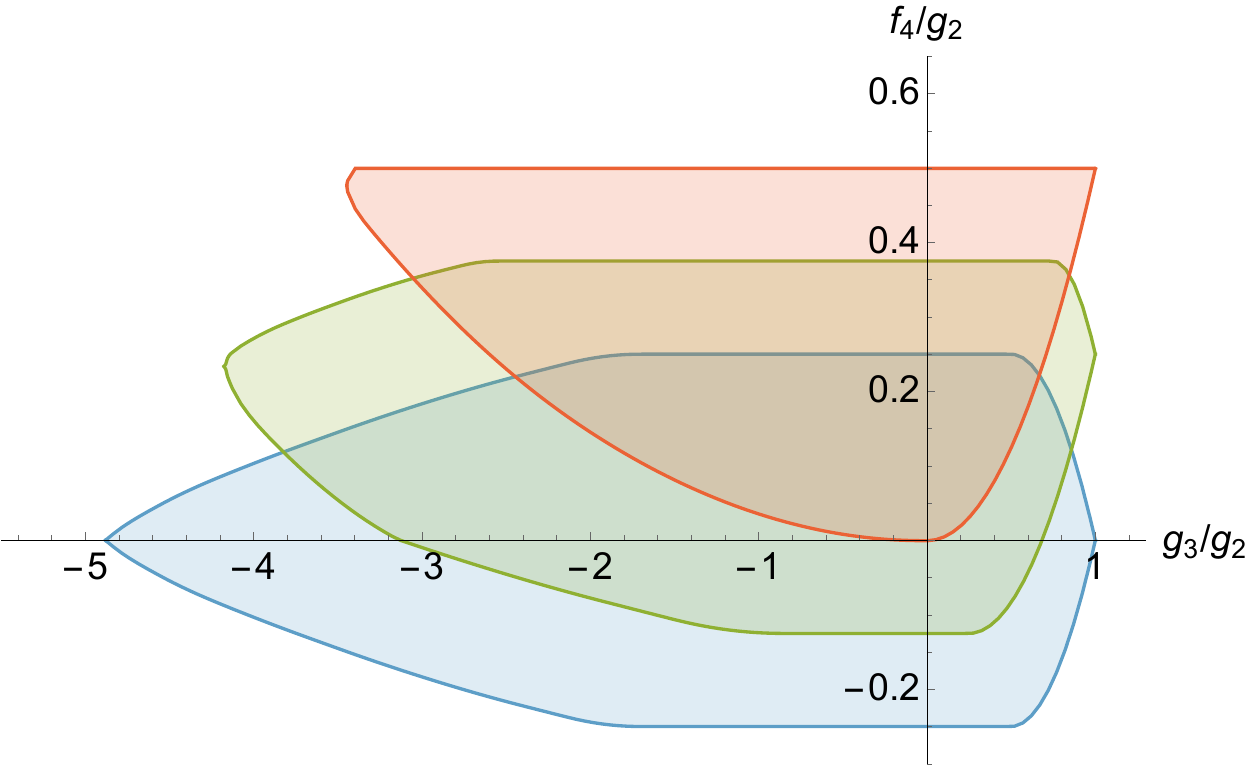}
\caption{Allowed region in the plane $(g_3/g_2, f_4/g_2)$ for fixed values of $f_2/g_2=k$. The blue, green, red region correspond respectively to $k=0,0.5,1$. Intermediate values can be obtained by linear interpolation.}\label{fig:g3f4}
\end{figure}

Using the same strategy, we find a numerical bound in the space $(g_3/g_2,\,f_4/g_2)$ and study how it changes as we vary $f_2$. The result is displayed in figure~\ref{fig:g3f4}.

Notice that for $f_2=g_2$, we have $f_4=\frac12(g_{4,1}+2g_{4,2})$, hence the two red regions in figures~\ref{fig:g3g4} and~\ref{fig:g3f4} have the same shape. 

\subsection{Comparison with partial UV completions}
\label{sec:partialUV}

In the sections above, we computed rigorous bounds in the EFT coefficients appearing in the low-energy amplitudes \eqref{eq:fgh}. It may be interesting to compare our bounds with values in some known partial UV completions. We use the word ``partial'' to emphasize that the theories we have in mind may not themselves be UV-complete theories, but they may be defined in an arbitrarily large energy range $M^2<E^2<\Lambda^2$, where $\Lambda$ is a scale at which further UV completion is necessary.

The partial completions we consider consist of integrating out massive fields, either at tree-level or one-loop level. In tables~\ref{tab:EFTcoefValues_trees} and \ref{tab:EFTcoefValues_loops} we give the values of the EFT coefficients following from this procedure. Note that for the case of fields with mass $m$ entering at loop level, we have $M^2=4m^2$. Any linear combination of the coefficients in tables~\ref{tab:EFTcoefValues_trees}--\ref{tab:EFTcoefValues_loops} will also be allowed.

\subsubsection{Tree-level completions}
\label{sec:treelevel}
\begin{table}[ht]
\centering
\caption{EFT coefficients that result from integrating out various particles of mass $M$ at tree-level.} \label{tab:EFTcoefValues_trees}
{
\renewcommand{\arraystretch}{1.5}
\begin{tabular}{|c|cc|ccc|ccc|}
\hline
 &  \multicolumn{2}{c|}{ $\Delta=8$} & \multicolumn{3}{c|}{ $\Delta=10$}& \multicolumn{3}{c|}{ $\Delta=12$} 
\\
Completion& $f_2$ & $g_2$ & $f_3$ & $g_3$ & $h_3$ & $f_4$ & $g_{4,1}$ & $g_{4,2}$ 

\\\hline\hline
Scalar
&  $\frac{4g^2}{M^4}$&$\frac{4g^2}{M^4}$
& $\frac{12g^2}{M^6}$ & $\frac{4g^2}{M^6}$ & $0$
& $\frac{2g^2}{M^8}$ & $\frac{4g^2}{M^8}$ & $0$

\\\hline
Axion
&  $-\frac{4g^2}{M^4}$&$\frac{4g^2}{M^4}$
& $\!-\frac{12g^2}{M^6}$ & $\frac{4g^2}{M^6}$ & $0$
& $\!-\frac{2g^2}{M^8}$ & $\frac{4g^2}{M^8}$ & $0$

\\\hline
Graviton
&  $0$
&
$\frac{8 g^2}{M^4}$
& $0$ & $\!-\frac{4 g^2}{ M^6}$ & $0$
& $0$ & $\!-\frac{4 g^2}{M^8}$ & $\frac{4 g^2}{M^8}$
\\\hline
\end{tabular}
}
\end{table}
First we consider partial completions which arise from integrating out a single particle at tree-level.

\paragraph{Scalar:}
Consider the action
\begin{align}
    \mathcal{L}_s = -\frac{1}{4} (F_{\mu \nu} F^{\mu \nu}) -\frac{1}{2} (\partial \phi)^2 - \frac{1}{2} M^2 \phi^2 + \frac{g}{M} \phi (F_{\mu \nu} F^{\mu \nu}). 
\end{align}
The equation of motion for $\phi$ reads
\begin{align}
    (M^2 - \Box) \phi = \frac{g}{M} (FF) \, .
\end{align}
Plugging that back into the Lagrangian gives
\begin{align}
\begin{split}
     \tilde{\mathcal{L}}_s &= -\frac{1}{4} (F_{\mu \nu} F^{\mu \nu}) + \frac{g^2}{2 M^4} FF \frac{1}{1 - \frac{\Box}{M^2}} FF\\
    & = -\frac{1}{4} (F_{\mu \nu} F^{\mu \nu})  + \frac{g^2}{2 M^4} (FF) \left[ 1 + \frac{\Box}{M^2} + \frac{\Box^2}{M^4} + \ldots \right]  (FF).
\end{split}
\end{align}
Computing the two-to-two photon amplitudes from $\mathcal{L}$ and $\tilde{\mathcal{L}}$ results in the same set of EFT coefficients, which are recorded in the first line of table~\ref{tab:EFTcoefValues_trees}. 

\paragraph{Axion:}
Starting from the action
\begin{align}
    \mathcal{L}_a = -\frac{1}{4} (F_{\mu \nu} F^{\mu \nu}) -\frac{1}{2} (\partial \phi)^2 - \frac{1}{2} M^2 \phi^2 + \frac{g}{M} \chi (F_{\mu \nu} \tilde{F}^{\mu \nu}),
\end{align}
the low-energy Lagrangian is identical to the scalar with the replacement $FF \to F \tilde F$:

\begin{align}
    \tilde{\mathcal{L}}_a  = -\frac{1}{4} (F_{\mu \nu} F^{\mu \nu})  + \frac{g^2}{2 M^4} (F \tilde F) \left[ 1 + \frac{\Box}{M^2} + \frac{\Box^2}{M^4} + \ldots \right]  (F \tilde F).
\end{align}
The values are reported in table~\ref{tab:EFTcoefValues_trees}.

\paragraph{Vector:}
One can ask if a massive vector provides a similar example of a partial completion. The answer is no.\footnote{We thank Callum Jones for helping to clarify this point.}

Consider a massive vector $V$ with field strength $G$ and coupled to the photon through a $VAA$ three-point interaction. One possible choice is
\begin{align}
    \mathcal{L}_v = -\frac{1}{4} (F_{\mu \nu} F^{\mu \nu}) -\frac{1}{4} (G_{\mu \nu} G^{\mu \nu}) - \frac{1}{2} M^2 V_\mu V^\nu + \frac{g}{M^2} (\partial_\rho V^\rho) (F_{\mu \nu} F^{\mu \nu}).
\end{align}
However, the amplitudes in this theory do not contain a pole at mass $M$. This can be understood by noting that the interaction is removable by a field redefinition:
\begin{align}
    V_\mu \to V_\mu - \frac{g}{M^4} \partial_\mu (F^2),
\end{align}
which results in the following Lagrangian
\begin{align}
    \tilde{\mathcal{L}}_v = -\frac{1}{4} (F_{\mu \nu} F^{\mu \nu}) -\frac{1}{4} (G_{\mu \nu} G^{\mu \nu}) - \frac{1}{2} M^2 V_\mu V^\nu + \frac{g^2}{2M^6} (\partial_\alpha F^2) (\partial^\alpha F^2).
\end{align}
Hence we see that the theory, through the field redefinitions, is equivalent to a non-interacting massive vector plus a single six-derivative interaction. This leads to amplitudes whose behavior is $\sim s^3$ at large $s$, in violation of our assumptions.

One might wonder if a different interaction, perhaps with more derivatives, could lead to an interacting partial completion. In fact, this is impossible on general grounds due to the Landau--Yang theorem \cite{Landau:1948kw,Yang1950}: angular momentum selection rules prevent a massive vector from decaying into two photons.

\paragraph{Graviton:}
For a massive tensor (see \cite{deRham:2014zqa} for a review) coupled to photons we have the Lagrangian
\begin{align}
    \mathcal{L}_t  = -\frac{1}{4} (F_{\mu \nu} F^{\mu \nu}) - \frac{1}{4} h^{\mu \nu} \mathcal{E}^{\alpha \beta}_{\mu \nu} h_{\alpha \beta} - \frac{M^2}{8}\left(h_{\mu \nu} h^{\mu \nu} - h^2 \right) + h_{\mu \nu} T^{\mu \nu} ,
\end{align}
where the kinetic operator is given by 
\begin{align}
    \mathcal{E}^{\alpha \beta}_{\mu \nu} h_{\alpha \beta}  = - \frac{1}{2} \left[  \Box h_{\mu \nu} - \partial_\mu \partial_\alpha h^\alpha_\nu - \partial_\nu \partial_\alpha h^\alpha_\mu +  \partial_\mu \partial_\nu h - \eta_{\mu \nu} \Box h + \eta_{\mu \nu} \partial^\alpha \partial^\beta h_{\alpha \beta} \right] .
\end{align}
In general, we could consider general couplings, where the tensor $T$ takes the form
\begin{align}
\begin{split}    
    & T_{\alpha \beta} =  \frac{g}{M}  F_{\alpha \delta} F_{\beta}{}^\delta + \frac{\tilde g}{M}  \delta_{\alpha \beta} F_{\mu \nu} F^{\mu \nu} .
\end{split}
\end{align}
The amplitudes in this theory grow like $s^3$ at large $s$. However, the combination $\tilde g = -\frac14 g$ only grows like $s^2$, so this is the combination we will consider here.\footnote{Technically we need it to fall off faster than $s^2$. The massless graviton also gives $s^2$ fall-off, but string theory softens the high-energy behavior of the graviton to $\mathcal{O}(s^{2 + \alpha' u})$, with $u<0$ in the physical scattering region. So $s^2$ may obey the bound at infinity, depending on the sign of the corrections, while $s^3$ has no chance of satisfying the required bound.} \footnote{Only one combination can give the correct high-energy behavior. To see this, consider the field redefinition
\begin{align}
    h_{\mu \nu} \to h_{\mu \nu} - \frac{4 g}{3M^3} \left( \eta_{\mu \nu} + \frac{2}{M^2} \partial_\mu \partial_\nu \right) F^2 \, .
\end{align}
This removes the interaction $\tilde g h FF$ but adds a four- and six-derivative self-interaction for $FF$. Therefore the $\tilde g$ coupling can be changed by tuning the four- and six-derivative interactions. However, these interactions give terms in the amplitude that fall off as $s^3$, so any deviation from the massive gravity coupling will fall off like $s^3$.}
For this particular choice, the tensor $T$ then becomes the canonical energy-momentum tensor of $F$, given by 
\begin{equation}
    T_{\mu \nu} = F_{\mu \alpha} F_\nu{}^\alpha - \frac{1}{4} F_{\alpha \beta} F^{\alpha \beta} \eta_{\mu \nu} \, . 
\end{equation}
The equations of motion simplify drastically in this case because the energy-momentum tensor is both conserved and traceless. As a result, we can write down the low-energy Lagrangian explicitly:
\begin{align}
\begin{split}
    \tilde{\mathcal{L}}_g &=  -\frac{1}{4} (F_{\mu \nu} F^{\mu \nu}) -2 T^{\mu \nu} \frac{1}{\Box - M^2} T_{\mu \nu} \\
    & =  -\frac{1}{4} (F_{\mu \nu} F^{\mu \nu}) + \frac{2 g^2}{M^4} \Big( F_{\mu \alpha} F_\nu{}^\alpha \left[ 1 + \frac{\Box}{M^2} + \frac{\Box^2}{M^4} + \ldots \right]  F^\mu{}_{\beta} F^{\nu \beta} \\
    & \qquad \qquad \qquad \qquad \qquad \qquad - \frac{1}{4} (FF) \left[ 1 + \frac{\Box}{M^2} + \frac{\Box^2}{M^4} + \ldots \right]  (FF) \Big) \, .
\end{split}
\end{align}
The corresponding Wilson coefficients are included in table~\ref{tab:EFTcoefValues_trees}. Let us stress that these values are speculative, since the graviton amplitudes $\mathcal{A}(s,t)$ grow like $s^2$ at fixed $u$, and therefore marginally violate the Froissart bound.

\subsubsection{Loop-level completions}
QED provides an important example of a loop-level completion. In this case, the higher-derivative coefficients can be extracted from the full one-loop amplitude first computed by Karplus and Neuman \cite{Karplus:1950zz}. 
\begin{table}[ht]
\centering
\caption{EFT coefficients in loop-level partial UV completions. Here $M^2 = 4 m_i^2$. Recall from \eqref{eq:a1a2g2f2refintro} that $a_1=\frac{g_2+f_2}{16}$, $a_2=\frac{g_2-f_2}{16}$.}\label{tab:EFTcoefValues_loops}
{
\renewcommand{\arraystretch}{1.5}
\begin{tabular}{|c|cc|ccc|ccc|}
\hline
 &  \multicolumn{2}{c|}{ $\Delta=8$} & \multicolumn{3}{c|}{ $\Delta=10$}& \multicolumn{3}{c|}{ $\Delta=12$} 
\\
Completion& $f_2$ & $g_2$ & $f_3$ & $g_3$ & $h_3$ & $f_4$ & $g_{4,1}$ & $g_{4,2}$ 
\\\hline\hline
QED 
& $-\frac{\alpha^2}{15m_e^4}$  & $\frac{11\alpha^2}{45m_e^4}$
& $\!-\frac{2\alpha^2}{63m_e^6}$ & $\frac{4\alpha^2}{315m_e^6}$ & $-\frac{\alpha^2}{315m_e^6}$
& $\!-\frac{\alpha^2}{945m_e^8}$ & $\frac{41\alpha^2}{18900m_e^8}$ & $\frac{\alpha^2}{756m_e^8}$ 
\\
& \multicolumn{2}{r|}{\cite{Euler:1936oxn,Euler:1935zz}}
& \multicolumn{3}{r|}{\cite{Karplus:1950zz,Costantini:1971cj}} 
& \multicolumn{3}{r|}{\cite{Karplus:1950zz,Costantini:1971cj}}
\\\hline
Scalar QED & $\frac{\tilde\alpha^2}{30m_{\tilde e}^4}$&$\frac{2\tilde\alpha^2}{45m_{\tilde e}^4}$
& $\frac{\tilde\alpha^2}{63m_{\tilde e}^6}$& $\frac{\tilde\alpha^2}{210m_{\tilde e}^6}$& $\frac{\tilde\alpha^2}{630m_{\tilde e}^6}$
& $\frac{\tilde\alpha^2}{1890m_{\tilde e}^8}$& $\!\frac{17\tilde\alpha^2}{18900m_{\tilde e}^8}\!$& $\frac{\tilde\alpha^2}{7560m_{\tilde e}^8}$ 
\\
& \multicolumn{2}{r|}{\cite{Weisskopf:1936hya}}
& \multicolumn{3}{r|}{\cite{Yang:1994nu}} 
& \multicolumn{3}{r|}{\cite{Yang:1994nu}}
\\\hline
${W^\pm}$ sector & $\frac{\alpha^2}{10m_W^4}$&$\frac{14\alpha^2}{5m_W^4}$ & $\frac{\alpha^2}{21m_W^6}$ & $\!\!-\frac{47\alpha^2}{630m_W^6}$ & $\frac{\alpha^2}{210m_W^6}$ & $\frac{\alpha^2}{630m_W^8}$ & $\!\!-\frac{83\alpha^2}{6300m_W^8}$ & $\frac{23\alpha^2}{840m_W^8}$ 
\\
 & \multicolumn{2}{r|}{\cite{Vanyashin:1965ple,Yang:1994nu}}& \multicolumn{3}{r|}{\cite{Yang:1994nu}}& \multicolumn{3}{r|}{\cite{Yang:1994nu}}
\\\hline
\end{tabular}
}
\end{table}
We extracted the coefficients from the expressions given in \cite{Costantini:1971cj}, who report some misprints in earlier literature. 
The relevant diagram to evaluate is a single electron $e$ running in a loop coupled to the external photons in the standard three-point interaction.

Likewise we may consider scalar QED and vector QED where a massive charged scalar/vector runs in the loop, coupling to the external photons with both three- and four-point interactions -- for the relevant diagrams see \emph{e.g.}\ \cite{Preucil:2017wen}. 
The values for $f_2$ and $g_2$ in vector QED agree with the phenomenologically more interesting case of a gauged $W$ boson, ``$W^\pm$ sector'', and for this case, as well as scalar QED, the complete one-loop four-point amplitudes were computed in \cite{Yang:1994nu}.


The bounds we provide in this paper are valid when the theory is infinitely weakly coupled at the scale $M$, but this is not the case for loop-level completions. The two-loop corrections to QED were computed by Ritus in the case of QED \cite{Ritus:1975cf} (or \cite{Ritus:1998jm})
\begin{equation}
f_2=\frac{\alpha^2}{15m_e^4}\left(
-1-\frac{25\alpha}{4\pi}
\right),\qquad 
g_2=\frac{11\alpha^2}{45m_e^4}\left(
1+\frac{1955\alpha}{396\pi}
\right) \, ,
\end{equation}
and scalar QED \cite{Ritus:1977iu}
\begin{equation}
f_2=\frac{\tilde\alpha^2}{30m_{\tilde e}^4}\left(1+\frac{45\tilde\alpha}{8\pi}\right)
,\qquad
g_2=\frac{2\tilde\alpha^2}{45m_{\tilde e}^4}\left(1+\frac{1535\tilde\alpha}{288\pi}\right).
\end{equation}
Using $\alpha\approx \frac1{137}$, the corrections in QED are numerically in the order $10^{-2}$. For our purposes, however, we may consider QED as a partial UV completion in the limit of arbitrarily weak coupling $\alpha$.

\section{Discussion}
In this paper, we have derived a number of two-sided bounds on EFT coefficients appearing in the effective low-energy theory of photons. Our method assumes only that the high-energy amplitudes are analytic on the upper half-plane, bounded at large $s$, and weakly coupled up to and including the scale $s \sim M^2$. To derive two-sided bounds on the ratios of our coefficients, we have incorporated the use of null constraints that were introduced in \cite{Caron-Huot:2020cmc} for the case of scattering massless scalars. 

For one potential application of our bounds, consider a theory with a non-minimally coupled scalar:
\begin{align}
    \mathcal{L} = -\frac{1}{4} F^2 -\frac{1}{2} (\partial \phi)^2 - \frac{1}{2} M^2 \phi^2 + \frac{k_1}{M} \phi F^2 + \frac{k_2}{M} (\Box \phi) F^2 + \ldots \, ,
\end{align}
where the $\ldots$ refer to further interactions, which must all have more than four derivatives. If we compute the amplitudes in this theory, we find 
\begin{align}
    g_2 = \frac{4 k_1^2}{M^4}, \qquad g_3 = \frac{4(k_1^2 + 2 k_1 k_2)}{M^6} \, .
\end{align}
The higher interactions cannot affect $g_2$ or $g_3$, so we directly conclude that $k_1 k_2\leqslant0$. This can be seen by computing the ratio, $M^2 g_3 / g_2 = 1 + 2 k_1 k_2  / k_1^2 $. Our bounds imply that this ratio must be less than one. Therefore our results, combined with the fact that the massive scalar lives at the boundary of our allowed region, constrain higher-derivative interactions in the massive scalar partial completion.

We have compared our bounds to a number of ``partial UV completions'' where the photons interact with only a single massive particle. The result was that the massive scalar and massive axion, which coupled to the photon at tree-level, saturated a number of the bounds. The massive graviton and the loop-level completions lie well inside our bounds. The bounds in section~\ref{sec:results} display a few other corners that are not populated by known theories. It would be very interesting if physical theories which live at these kinks could be identified -- perhaps theories arising from dimensional reduction or braneworld scenarios could be potential candidates. Or perhaps they are not physical, in which case one should ask what assumptions about the $S$-matrix are required to rule out the amplitudes saturating these bounds. 

We also pointed that the putative tree-level completion involving a massive vector is trivial because the interactions may be removed by field redefinitions. This is also the case when considering a graviton with only $h_{\mu}^\mu F^2$ coupling, as well as the ``axi-vector'' and ``axi-tensor,'' which couple to $F \tilde F$. In general, these theories may violate our bounds. This is related to the argument in \cite{Hamada:2018dde}, where it is shown that such a trivial theory -- a graviton with only $h_\mu^\mu F^2$ coupling -- violates the conjecture of \cite{Cheung:2018cwt} that higher-derivative corrections always increase the Wald entropy of thermodynamically stable black holes. Clearly, such a theory cannot give rise to a unitary $S$-matrix -- the point of \cite{Hamada:2018dde} was to highlight that the behavior of the $S$-matrix is a clean, field-redefinition invariant way to diagnose which theories should be taken seriously when making Swampland-like arguments.

There are a number of directions that deserve to be further explored. A crucial one is to understand the effect of loops. The assumption of weak coupling at the scale $M$ means that loop effects of the high-energy theory may be safely ignored, but it limits the scope of the bootstrap method. Because of this assumption, we are only able to use part of the unitarity constraints -- that some of $\rho_i \geqslant 0$ -- and not the full constraints, which also bound $\rho_i$ from above, because such an upper bound is trivially satisfied at weak coupling, where $\rho_i \ll 1$.

One consequence of the weak coupling assumption is that $\rho_5$ does not participate in any of the unitarity inequalities we derived in section~\ref{sec:three}. As a result, we are unable to place any bounds on the coefficients $h_i$. It is interesting that these coefficients are absent in all of the tree-level completions that we studied. We believe a more general unitarity set-up could allow us to bound $\rho_5$ as well.

Another important direction is to extend the considerations to photons coupled to gravity. The appropriate parametrization of the low-energy amplitudes in that case was outlined in \cite{Cheung:2014ega}. This has the benefit that it could shed light on the weak gravity conjecture \cite{Arkani-Hamed:2006emk,Kats:2006xp}. Previous discussions on the use of positivity bounds to prove the WGC include \cite{Hamada:2018dde,Bellazzini:2019xts,Alberte:2020jsk}; so far the so-called $t$-channel pole appearing from in the amplitude due to graviton exchange has been a major obstacle to a convincing proof. Recently it was shown in \cite{Caron-Huot:2021rmr} that this obstacle can be circumvented in the case of scalar scattering by taking an integral of the subtracted amplitude centered around small $t$, rather than the $t \to 0$ limit. This method is limited to $d > 4$, so pursuing this line requires extending our results to $d > 4$, and/or extending the methods of \cite{Caron-Huot:2021rmr} to $d = 4$.

\appendix

\section{Unitarity}
\label{app:unitarity}

Unitarity requires that all physical states have positive norm. If we have a basis of states labeled by $i$, $j$, then unitarity may be expressed by the requirement $\langle i | j \rangle \succeq 0$, \emph{i.e.} that matrix of inner products is positive semi-definite. 

We are concerned with photon scattering. The unitarity constraints on amplitudes with spinning external states was nicely reviewed in \cite{Hebbar:2020ukp}; our discussion will largely parallel theirs. The basis of states for two-to-two photon scattering is the set of incoming and outgoing two-particle states which transform in irreducible representations of the Poincar{\'e} group. Such states are labeled by squared COM energy $s$, momentum $\vec{p}$, spin $J$, and helicity $\lambda$, as well as the helicities of each constituent particle, $\lambda_1$ and $\lambda_2$, which equal $\pm 1$. Because the particles are indistinguishable, we have $|+-\rangle = |-+\rangle$. So we need consider the following basis of states:
\begin{align}\nonumber
    & |a\rangle_{\text{in}}  = |s, \vec{p}, J, \lambda, +, + \rangle_{\text{in}} \, ,&& |b\rangle_{\text{in}} =  |s, \vec{p}, J, \lambda, +, - \rangle_{\text{in}} \, , &&  |c\rangle_{\text{in}} =  |s, \vec{p}, J, \lambda, -, - \rangle_{\text{in}} \, , \\
    & |a\rangle_{\text{out}}  = |s, \vec{p}, J, \lambda, +, + \rangle_{\text{out}} \, ,&& |b\rangle_{\text{out}} =  |s, \vec{p}, J, \lambda, +, - \rangle_{\text{out}} \, , && |c\rangle_{\text{out}} =  |s, \vec{p}, J, \lambda, -, - \rangle_{\text{out}} \, .
\end{align}
The states $|a\rangle_i$ and $|c\rangle_i$ exist only for even spins $J$, while $|b\rangle_i$ exists for all spins. There is no mixing between different spin eigenstates, so the positive semi-definiteness of the entire matrix of states will imply positive semi-definiteness on the matrix of states at each spin. Furthermore, we have assumed parity invariance so there is no mixing between even and odd parity states. Thus it is convenient to define odd parity eigenstates
\begin{align}
    |1\rangle_i := \frac{1}{\sqrt{2}} \,  (|a\rangle_i - |c\rangle_i ), \qquad J = 0, 2, 4, \ldots
\end{align}
and even parity eigenstates
\begin{align}
\begin{split}
        &|2\rangle_i := \frac{1}{\sqrt{2}} \, (|a\rangle_i + |c\rangle_i ), \qquad J= 0, 2, 4, \ldots \\
        &|3\rangle_i := \sqrt{2} \,  |b\rangle_i \, , \qquad \qquad \qquad \, J =  2, 3, 4, \ldots \, .
\end{split}
\end{align}
Note that in last line $|3\rangle_i$ is only defined for $J\geq2$. This is because the two-particle state formed from incoming particles with helicity $\lambda_1$ and $\lambda_2$ is $\lambda = \lambda_1-\lambda_2$. Therefore $|3\rangle_i$ has helicity $\lambda = 2$. Helicity $2$ is only a possible eigenvalue for angular momentum $J \geq 2$.

Now we can look at the matrix of inner products of all ingoing and outgoing states, and study the corresponding positivity conditions. For the parity odd sector, we find that unitarity implies
\begin{align}
    \begin{pmatrix}
    {}_\text{in} \langle 1 | 1 \rangle_{\text{in}} && {}_\text{in} \langle 1 | 1 \rangle_{\text{out}} \\
    {}_\text{out} \langle 1 | 1 \rangle_{\text{in}} && {}_\text{out} \langle 1 | 1 \rangle_{\text{out}}
    \end{pmatrix} \succeq 0 \, ,\qquad J = 0, 2, 4, \ldots.
\end{align}
For parity even states, we have
\begin{align}
    \begin{pmatrix}
    {}_\text{in} \langle 2 | 2 \rangle_{\text{in}} && {}_\text{in} \langle 2 | 2 \rangle_{\text{out}} \\
    {}_\text{out} \langle 2 | 2 \rangle_{\text{in}} && {}_\text{out} \langle 2 | 2 \rangle_{\text{out}}
    \end{pmatrix} & \succeq 0 \, , \qquad J = 0, \\
    \begin{pmatrix}
    {}_\text{in} \langle 3 | 3 \rangle_{\text{in}} && {}_\text{in} \langle 3 | 3 \rangle_{\text{out}} \\
    {}_\text{out} \langle 3 | 3 \rangle_{\text{in}} && {}_\text{out} \langle 3 | 3 \rangle_{\text{out}}
    \end{pmatrix} & \succeq 0  \, , \qquad J =  3, 5, 7, \ldots.
\end{align}
Finally, for even spins greater than $0$, the $|2 \rangle$ and $|3 \rangle$ even parity states can mix, leading to the larger matrix
\begin{align}
    \begin{pmatrix}
    {}_\text{in} \langle 2 | 2 \rangle_{\text{in}} && {}_\text{in} \langle 2 | 3 \rangle_{\text{in}} && {}_\text{in} \langle 2 | 2 \rangle_{\text{out}} && {}_\text{in} \langle 2 | 3 \rangle_{\text{out}} \\ 
    {}_\text{in} \langle 3 | 2 \rangle_{\text{in}} && {}_\text{in} \langle 3 | 3 \rangle_{\text{in}} && {}_\text{in} \langle 3 | 2 \rangle_{\text{out}} && {}_\text{in} \langle 3 | 3 \rangle_{\text{out}} \\ 
    {}_\text{out} \langle 2 | 2 \rangle_{\text{in}} && {}_\text{out} \langle 2 | 3 \rangle_{\text{in}} && {}_\text{out} \langle 2 | 2 \rangle_{\text{out}} && {}_\text{out} \langle 2 | 3 \rangle_{\text{out}} \\ 
    {}_\text{out} \langle 3 | 2 \rangle_{\text{in}} && {}_\text{out} \langle 3 | 3 \rangle_{\text{in}} && {}_\text{out} \langle 3 | 2 \rangle_{\text{out}} && {}_\text{out} \langle 3 | 3 \rangle_{\text{out}} 
    \end{pmatrix}  & \succeq 0 \, , \qquad J = 2, 4, 6, \ldots.
\end{align}

To derive positivity constraints from this, we need 
\begin{align}
    {}_\text{in} \langle i | j \rangle_{\text{in}}  = {}_\text{out} \langle i | j \rangle_{\text{out}} = \delta_{ij}\, , \qquad {}_\text{out} \langle i | j \rangle_{\text{in}} = \delta_{ij} + i T_{ij}\,,
\end{align}
where $T_{ij}$ represents the interacting part of the $S$-matrix. 
Using this, and applying positive semi-definiteness to each spin individually, we find the following conditions on the partial waves:
\begin{align}\label{eq:ineq1}
    2 \IM[A^2_J - A^1_J] & \ \geqslant \ |A^2_J - A^1_J|^2 \ \geqslant \ 0 \, , & J &= 0, 2, 4, \ldots, \\
    2 \IM[A^2_J + A^1_J] & \ \geqslant \  |A^2_J + A^1_J|^2 \ \geqslant \ 0 \, , & J &= 0, \\
    \IM[A^3_J] & \ \geqslant \  |A^3_J|^2 \ \geqslant \ 0 \, ,  & J &= 3, 5,7, \ldots.
\end{align}
These conditions arise from positivity of the three two-by-two matrices above. From examining the $2 \times 2$ principal minors of the larger matrix, we find
\begin{align}
    \begin{split}
    2 \IM[A^2_J + A^1_J] & \ \geqslant \  |A^2_J + A^1_J|^2 \ \geqslant \ 0, \qquad  \\
    \IM[A^3_J] & \ \geqslant \  |A^3_J|^2 \ \geqslant \ 0  \, , \qquad \quad \qquad J = 2, 4, 6, \ldots .\\
    1 &  \ \geqslant \ 4 |A^5_J |^2 \geqslant 0\, ,
    \end{split}
    \label{eq:ineq4}
\end{align}
Positivity of the three-by-three principal minors does not lead to any new positivity conditions.\footnote{It does, however, lead to different conditions such as $2 \IM[A^2_J + A^1_J]  \ \geqslant\  |A^2_J + A^1_J|^2 + 4 |A^5_J |^2  \ \geqslant \ 0$, which is strictly stronger than the above, but no more useful for our purposes.}

These equations give us positive partial wave expansions for three of our five amplitudes. However, the $t$-$u$ crossing symmetry which relates $\mathcal{A}^{+--+}$ and $\mathcal{A}^{+-+-}$ leads to another relation:\footnote{This is derived by relating the two partial wave expansions and using $d_{a, b}(\theta) = d_{a, -b}(\pi - \theta)$.}
\begin{align}
    A^4_J = (-1)^J A^3_J.
\end{align}
To summarize our positivity properties, we first define the spectral densities $\rho^i_J = \IM A^i_J$. These satisfy
\begin{align}
    \rho^2_J \pm \rho^1_J \ \geqslant \ 0, \qquad J = 0, 2, 4, \ldots, \\
    \rho^3_J \ \geqslant \ 0, \qquad J = 2, 3, 4, \ldots,
\end{align}
where we use the weak coupling assumption to simplify the inequalities \eqref{eq:ineq1}--\eqref{eq:ineq4}.
These conditions will allow us to derive sum rules which bound the EFT coefficients defined in \eqref{eq:fgh}.

\section{Details on analytic bounds}

\subsection{Simple analytic bound using one null constraint}
\label{app:simpleanalyicbound}

Here we will derive a two-sided bound on $g_3/g_2$ using the problem defined by \eqref{eq:normalizationrulemain}--\eqref{eq:nullWmain}.
It is instructive to think of these sum rules in terms of a reduced bracket, defined by including the positive quantity $1/(m^4g_2)$ in the integration measure:
\begin{equation}\label{eq:reducedbracketdef}
    \Big\langle\cdots\Big\rangle_{\mathrm{red}., i}:=\left\langle\frac1{m^4g_2}(\cdots)\right\rangle_i, \qquad i=0,e,o.
\end{equation}
Using this definition we can write the sum rules \eqref{eq:normalizationrulemain}--\eqref{eq:nullWmain} as
\begin{align}
\label{eq:normalizationrule}
    1&=\langle1\rangle_{\mathrm{red.},0}+\langle1\rangle_{\mathrm{red.},e}+\langle1\rangle_{\mathrm{red.},o}\,,
    \\
    \label{eq:g3sumrule}
    \frac{g_3}{g_2}&=\big\langle W^0_{g_3}\big\rangle_{\mathrm{red}.,0}+\big\langle W^e_{g_3}\big\rangle_{\mathrm{red}.,e}+\big\langle W^o_{g_3}\big\rangle_{\mathrm{red}.,o}\,,
   \\
   \label{eq:nullW}
   0& =\big\langle W^0_{Y_1}\big\rangle_{\mathrm{red.},0}+\big\langle W^e_{Y_1}\big\rangle_{\mathrm{red.},e}+\big\langle W^o_{Y_1}\big\rangle_{\mathrm{red.},o}\,,
\end{align}
where 
\begin{align}
W^0_{g_3}&=\frac{3 - 2\mathcal J^2}{3m^2},& W^e_{g_3}&= \frac{11-2\mathcal J^2}{3m^2}, & W^o_{g_3}&= \frac{11-2\mathcal J^2}{3m^2},
   \\
W^0_{Y_1}&= \frac{\mathcal J^2(\mathcal J^2-8)}{ m^4},
& W^e_{Y_1}&= \frac{7 \mathcal{J}^4-98 \mathcal{J}^2+312}{6 m^4},
& W^o_{Y_1}&= \frac{5 \mathcal{J}^4-94 \mathcal{J}^2+312}{6 m^4}.
\label{eq:n1Wrules}
\end{align}

We will now derive the lower bound $r_{\mathrm{min}}$ defined by
\begin{equation}
\frac{r_{\mathrm{min}}}{M^2}\ \leqslant \ \frac{g_3}{g_2}.
\end{equation}
The system of sum rules \eqref{eq:normalizationrule}--\eqref{eq:nullW} can be interpreted in the following way. The right-hand side in each of these equations is a positive linear combination of the terms entering in the brackets. 
The first equation \eqref{eq:normalizationrule} determines the normalization, which means that the region generated by the expressions in the \eqref{eq:g3sumrule}--\eqref{eq:nullW} constitutes a convex hull of the points of the form $\vec W^i(m^2,J)=(W^i_{g_3}(m^2,J),W^i_{Y_1}(m^2,J))$ for $i=0,e,o$ in the $(e_1,e_2)$ plane. The intersection of this convex hull with the $e_1$ axis determines the range of the allowed values of $g_3/g_2$.

\begin{figure}[ht]
  \centering
\includegraphics[width=\textwidth]{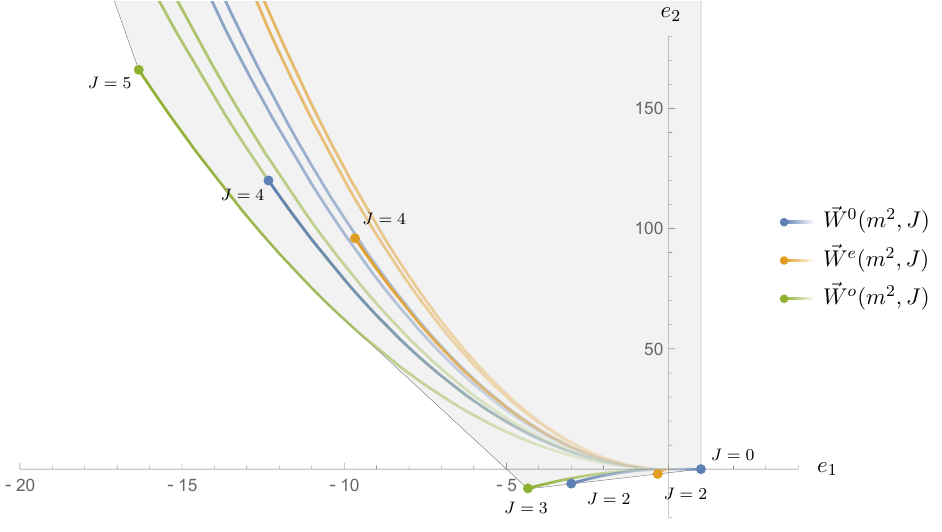}
\caption{Plot showing the convex hull of points in the right-hand side of equations~\eqref{eq:g3sumrule} and \eqref{eq:nullW}. The bound \eqref{eq:boundg3g2} is the intersection between the shaded region and the $e_1$ axis. The plot uses units where $M^2=1$.}\label{fig:simplebound}
\end{figure}

In figure~\ref{fig:simplebound} we plot the vectors $\vec W^i$ for $J\leqslant9$ and $m^2\geqslant M^2$. In the $(e_1,e_2)$ plane, each spin $J$ determines a section of a parabola starting at $m^2=M^2$.
The convex hull of all the allowed points for all $J$ is indicated in gray.\footnote{We have excluded higher values of $J$ to avoid cluttering the figure. 
The curves corresponding to higher $J$ are always in the gray-shaded region and do not affect the bounds.} 
It is clear that the upper bound comes from the point $\vec W^0(M^2,0)$, giving immediately the bound \eqref{eq:upperboundg3}. From the figure, we can also see that the lowest possible value of $g_3/g_2$ must come from a linear combination of the point $\vec W^o(M^2,3)$ and a point on the curve $\vec W^{o}(m^2,5)$ parametrized by $m^2\geqslant M^2$. Optimizing over $m^2$ and computing the intersection with the $e_1$ axis gives
\begin{equation}
    r_{\mathrm{min}}=-\left(\frac{13}6+\sqrt{\frac{7877}{996}}\right),
\end{equation}
as quoted in \eqref{eq:boundg3g2} in the main text.

\subsubsection[Bound when $f_2=g_2$]{Bound when $\boldsymbol{f_2=g_2}$}
\label{app:f2equalsg2}

Note first that assuming $f_2=g_2$ implies that $g_3=3f_3$, since multiplying \eqref{eq:g3plusx1f3} by $g_2+x_1f_2$ and taking the limit $x_1\to-1$ shows that $0\leqslant g_3-\frac13f_3\leqslant0$. Likewise with $f_2=-g_2$ we must have $f_3=-3g_3$.

Using the $\pm$ brackets, we can write the normalization sum rule~\eqref{eq:normalizationrule} as
\begin{equation}\label{eq:fsumrule}
    1=\left\langle\tfrac12\right\rangle_{\mathrm{red.},+}+\left\langle\tfrac12\right\rangle_{\mathrm{red.},-}+\langle1\rangle_{\mathrm{red.},e}+\langle1\rangle_{\mathrm{red.},o}\,,
\end{equation}
Next, using $f_3=3g_3$, averaging the sum rules for $g_3$ and $\frac13f_3$ (see \eqref{eq:f3g3main}) we get
\begin{equation}
\frac{g_3}{g_2}=\big\langle \tfrac12W^0_{g_3}\big\rangle_{{\mathrm{red}}.,+}+\big\langle \tfrac12W^e_{g_3}\big\rangle_{{\mathrm{red}}.,e}+\big\langle \tfrac12 W^o_{g_3}\big\rangle_{{\mathrm{red}}.,o}\,,
\end{equation}
Likewise adding the $g$-type and $f$-type null constraint we get
\begin{equation}
0=\big\langle \tfrac12W^0_{Y_1}\big\rangle_{{\mathrm{red}}.,+}+\big\langle \tfrac12W^e_{Y_1}\big\rangle_{{\mathrm{red}}.,e}+\big\langle \tfrac12 W^o_{Y_1}\big\rangle_{{\mathrm{red}}.,o}\,,
\label{eq:Ynullpluseo}
\end{equation}
We can bring the problem defined by \eqref{eq:fsumrule}--\eqref{eq:Ynullpluseo} onto a better form by re-defining the reduced brackets in the $\pm$ case,
\begin{equation}\label{eq:reducedbracketpm}
    \Big\langle\cdots\Big\rangle_{\widehat{\mathrm{red}}., \pm}:=  \left\langle\frac12(\cdots)\right\rangle_{{\mathrm{red}}., \pm}\,,
\end{equation}
so that 
\begin{align}
    1&=\langle1\rangle_{\widehat{\mathrm{red}}.,+}+\langle1\rangle_{\mathrm{red.},-}+\langle1\rangle_{\mathrm{red.},e}+\langle1\rangle_{\mathrm{red.},o}\,,
    \\
\frac{g_3}{g_2}&=\big\langle W^0_{g_3}\big\rangle_{\widehat{\mathrm{red}}.,+}+\big\langle \tfrac12W^e_{g_3}\big\rangle_{
\mathrm{red}.,e}+\big\langle \tfrac12 W^o_{g_3}\big\rangle_{\mathrm{red}.,o}\,,
\\
0&=\big\langle  W^0_{Y_1}\big\rangle_{\widehat{\mathrm{red}}.,+}+\big\langle \tfrac12W^e_{Y_1}\big\rangle_{\mathrm{red}.,e}+\big\langle \tfrac12 W^o_{Y_1}\big\rangle_{\mathrm{red}.,o}\,,
\end{align}
Comparing with the previous problem defined by \eqref{eq:normalizationrule}--\eqref{eq:nullW}, we note that compared to figure~\ref{fig:simplebound} we should rescale the curves corresponding to $\vec W^{e}(m^2,J)$ and $\vec W^o(m^2,J)$ by a factor $\frac12$. The minimal value is now given by interpolating between the point $\vec W^{0}(1,2)$ and $\vec W^0(m^2,4)$. Optimizing over $m^2$ we get that
\begin{equation}
g_3\geqslant -\left(\frac32+\sqrt{\frac{2989}{720}}\right) = -3.537496, \qquad f_2=g_2.
\end{equation}
This bound agrees with ($\frac13$ times) the value $\kappa(4)$ found in the scalar case in \cite{Caron-Huot:2021rmr,Caron-Huot:2020cmc}.

\subsection{Triangular region in the eight-derivative coefficient space}
\label{app:triangle}

Here we will show how to determine the triangular region in the $(g_{4,1},\,g_{4,2})$ plane given by \eqref{eq:triangle} in the main text. We will again work with the reduced brackets defined in \eqref{eq:reducedbracketdef}, and use the sum rules
\begin{align}
 1&=\langle1\rangle_{\mathrm{red.},0}+\langle1\rangle_{\mathrm{red.},e}+\langle1\rangle_{\mathrm{red.},o}\,,
 \\
\frac{ g_{4,i}}{g_2}&=\big\langle W^0_{g_{4,i}}\big\rangle_{\mathrm{red},0}+\big\langle W^e_{g_{4,i}}\big\rangle_{\mathrm{red},e}+\big\langle W^o_{g_{4,i}}\big\rangle_{\mathrm{red},o}\,,
 \\
0&=\big\langle W^0_{Y_1}\big\rangle_{\mathrm{red.},0}+\big\langle W^e_{Y_1}\big\rangle_{\mathrm{red.},e}+\big\langle W^o_{Y_1}\big\rangle_{\mathrm{red.},o}\,,
\end{align}
where
\begin{align}
W_{g_{4,1}}^0&=\frac{1}{ m^4}, & 
W_{g_{4,1}}^e&=\frac{-\mathcal{J}^4+2 \mathcal{J}^2+12}{12 m^4}, &
w_{g_{4,1}}^o&=\frac{\mathcal{J}^4-2 \mathcal{J}^2+12}{12 m^4},
\label{eq:Vg41}
\\
W_{g_{4,2}}^0&=0,&
W_{g_{4,2}}^e&=\frac{\mathcal{J}^4-2\mathcal J^2}{24 m^4} ,&
W_{g_{4,2}}^o&=\frac{-\mathcal{J}^4+2\mathcal J^2}{24 m^4},
\end{align}
and $W^i_{Y_1}$ are given in \eqref{eq:n1Wrules}.

We first show that
\begin{align}\nonumber
\frac{g_{4,1}+2g_{4,2}}{g_2}&=\left\langle\frac1{m^4}\right\rangle_{\mathrm{red.},0}+\left\langle\frac1{m^4}\right\rangle_{\mathrm{red.},e}+\left\langle\frac1{m^4}\right\rangle_{\mathrm{red.},o}\\&\leqslant\frac1{M^4}\left(
\left\langle1\right\rangle_{\mathrm{red.},0}+\left\langle1\right\rangle_{\mathrm{red.},e}+\left\langle1\right\rangle_{\mathrm{red.},o}
\right)=\frac{1}{M^4},
\end{align}
by which we see that
\begin{equation}
\label{eq:upperg41g42}
g_{4,1}+2g_{4,2}\leqslant \frac{g_2}{M^4}.
\end{equation}

We then define
\begin{equation}
\vec W^b=(W_{g_{4,1}}^b,W_{g_{4,2}}^b,W_{Y_1}^b),
\end{equation}
and consider the following problem: 
\emph{For which $A$ is it possible to find a $c_1$ such that
\begin{equation}\label{eq:inequalityg41g42bound}
(1,A,c_1)\cdot \vec W^b(m^2,J)\geqslant0,\quad b=0,e,o,
\end{equation}
for all $m^2\geqslant1$ and $J$ in the sums defining the respective brackets?}\footnote{For $b=0$, $J=0,2,4,\ldots$; for $b=e$, $J=2,4,6,\ldots$; and for $b=o$, $J=3,5,7,\ldots$.}

By limiting to spins $J\leqslant J_{\mathrm{max}}$ for some finite $J_{\mathrm{max}}$, the inequality~\eqref{eq:inequalityg41g42bound} can be simplified to a set of inequalities in the $(A,c_1)$ plane. The upper and lower value for $A$ resulting from these inequalities does not change by increasing $J_{\mathrm{max}}$ as long as $J_{\mathrm{max}}\geqslant5$:
\begin{equation}
A_{\mathrm{min}}=\frac{11}9,\quad \text{at }c_1=\frac1{18},\qquad A_{\mathrm{max}} = \frac{399}{185},\quad\text{at } c_1=\frac1{74}.
\end{equation}
This means that we have found the bound
\begin{equation}\label{eq:analyticg41g42}
g_{4,1}+Ag_{4,2}\geqslant0,\quad \forall\ A\in[A_{\mathrm{min}},A_{\mathrm{max}}]=[1.222,\ 2.157].
\end{equation}

By finding the intersection with the bound \eqref{eq:upperg41g42} we have limited the allowed region in the $(g_{4,1},g_{4,2})$ plane to the triangle given by \eqref{eq:triangle} and shown in figure~\ref{fig:g41g42}.

\section{Details on numerical implementation}

\subsection{Sum rules}
\label{app:appendixsumrules}

We consider the following system\footnote{For bounds that involve only the constants $g_{k(,i)}$, for which $\vec V^+=\vec V^-$, we will use that
\begin{equation*}
\big\langle \vec V\, \big\rangle_++\big\langle \vec V\,\big\rangle_-=\big\langle 2\vec V\,\big\rangle_0.
\end{equation*}
}
\begin{equation}
\sum c_i\vec C_{g_i} = \left\langle\vec V^+(m^2,J)\right\rangle_++\left\langle\vec V^-(m^2,J)\right\rangle_-+\left\langle\vec V^e(m^2,J)\right\rangle_e+\left\langle\vec V^o(m^2,J)\right\rangle_o\,.
\end{equation}
Here the sum goes over a range that covers a selection of size $N_c$ of the constants $c_i$, which include both the $g$ and $f$ EFT coefficients. The vectors $\vec V^b(m^2,J)$ are length $N_c+N_f+N_g$ vectors of the form
\begin{equation}
\vec V^b(m^2,J)=\big( V_{f_2}^b,V_{f_3}^b, V_{f_4}^b,V_{g_2}^b,V_{g_3}^b,V_{g_{4,1}}^b,V_{g_{4,2}}^b |X_1^b,\ldots,X^b_{N_{f}}|Y_1^b,\ldots,X^b_{N_{g}}\big),
\end{equation}
where each entry is a function of $m^2$ and $J$, and $b=+,-,e,o$. This corresponds to writing $\vec C_{g_i}$ in a diagonal basis. $N_f$ and $N_g$ denote the number of $f$-type and $g$-type null constraints used.

In table~\ref{tab:EFTbrackets} we collect the explicit form of the $V_{c_i}^b$ for the EFT coefficients considered in this paper. They were derived using the sum rules of section~\ref{sec:sumrules} in the main text. From eight-derivative order and onwards, they are unique only up to the addition of null constraints.

\begin{table}
\caption{Solving the sum rules gives the EFT coefficients in terms of brackets. Below are the terms which appear in each bracket in the solutions.}
\begin{center}
{
\renewcommand{\arraystretch}{1.5}
\begin{tabular}{ |c | c  c  c  c |}
\hline
  bracket $b$ & $+$ & $-$ & $e$ & $o$\\
  \hline \hline
  $V^b_{f_2}$ & $\frac{1}{2 m^4}$ & $- \frac{1}{2 m^4}$ & 0 & 0 \\
  \hline
  $V^b_{f_3}$ & $\frac{3-2\mathcal J^2}{2m^6}$ & $-\frac{3-2\mathcal J^2}{2m^6}$ & $0$ & $0$  \\
  \hline
  $V^b_{f_4}$ & $\frac{1}{4 m^8}$ & $-\frac{1}{4 m^8}$ & $0$ & $0$\\
  \hline
  $V^b_{g_2}$ & $\frac{1}{2 m^4} $ & $\frac{1}{2 m^4} $&  $\frac{1}{m^4} $ & $\frac{1}{m^4} $ \\
  \hline
  $V^b_{g_3}$ & $\frac{3-2\mathcal J^2}{6m^6}$ & $\frac{3-2\mathcal J^2}{6m^6}$ & $\frac{11-2\mathcal J^2}{3m^6}$  & $\frac{11-2\mathcal J^2}{3m^6}$  \\
  \hline
  $V^b_{g_{4,1}}$ & $\tfrac{1}{2 m^8}$ & $\tfrac{1}{2 m^8}$ & $\tfrac{-\mathcal{J}^4+2 \mathcal{J}^2+12}{12 m^8}$ & $\tfrac{\mathcal{J}^4-2 \mathcal{J}^2+12}{12 m^8}$\\
  \hline
  $V^b_{g_{4,2}}$ & $0$ & $0$ & $\tfrac{\mathcal{J}^2 (\mathcal{J}^2-2)}{24 m^8}$ & $-\tfrac{\mathcal{J}^2 (\mathcal{J}^2-2)}{24 m^8}$ \\
  \hline
\end{tabular}
}
\end{center}
\label{tab:EFTbrackets}
\end{table}
%

The null constraints are computed (up to an overall factor) using the replacement rules \eqref{eq:modifynullconstraintsf}--\eqref{eq:modifynullconstraintsf} given in the main text. 
The first two null constraints of each type are included in table~\ref{tab:Nullconstraintslist}. Note that we can always choose the form of the null constraints such that $X_i^+=-X_i^-=Y_i^+=Y_i^-$.

\begin{table}
\begin{center}
\caption{Solving the sum rules gives the EFT coefficients in terms of brackets. Below are the terms which appear in each bracket in the solutions.}
\label{tab:Nullconstraintslist}
{
\renewcommand{\arraystretch}{1.5}
\begin{tabular}{ |c | c  c  c  c |}
\hline
  bracket $b$ & $+$ & $-$ & $e$ & $o$\\
\hline\hline
$X_1^b$ 
&
$\frac{\mathcal J^2(\mathcal J^2-8)}{ m^8}$ & $-\frac{\mathcal J^2(\mathcal J^2-8)}{ m^8}$ 
& $0$ & $0$
\\\hline
$X_2^b$
&
$\frac{\mathcal{J}^2 (2 \mathcal{J}^4-43 \mathcal{J}^2+150)}{ m^{10}}$ & $-\frac{\mathcal{J}^2 (2 \mathcal{J}^4-43 \mathcal{J}^2+150)}{ m^{10}}$ 
& $0$ & $0$
\\\hline\hline
$Y_1^b$
&
$\frac{\mathcal J^2(\mathcal J^2-8)}{ m^8}$ & $
\frac{\mathcal J^2(\mathcal J^2-8)}{ m^8}$ & 
$\frac{7 \mathcal{J}^4-98 \mathcal{J}^2+312}{3 m^8}$ &
$\frac{5 \mathcal{J}^4-94 \mathcal{J}^2+312}{3 m^8}$
\\\hline
\multirow{2}{*}{$Y_2^b$}
& 
  \multirow{2}{*}{  $\frac{\mathcal{J}^2 (2 \mathcal{J}^4-43 \mathcal{J}^2+150)}{ m^{10}} $}&
 \multirow{2}{*}{  $\frac{\mathcal{J}^2 (2 \mathcal{J}^4-43 \mathcal{J}^2+150)}{ m^{10}} $}&
    $\frac{26 \mathcal{J}^6-763 \mathcal{J}^4}{5 m^{10}}\quad $&
    $\frac{14 \mathcal{J}^6-577 \mathcal{J}^4}{5 m^{10}}\quad$

\\
&&&     $+\frac{6462 \mathcal{J}^2-17640}{5 m^{10}} $&
    $+\frac{6138 \mathcal{J}^2-17640}{5 m^{10}}$
\\\hline
\end{tabular}
}
\end{center}
\end{table}

\subsection{Optimization problem}
\label{subsec:OptProblem}
Following the previous considerations in section~\ref{sec:three}, we have a relationship between each low-energy coefficient with a higher energy part of the form
\begin{equation}
g_i=\sum_{b}\left\langle V_{g_i}^{b}(m^2,J) \right\rangle_b\,,
\end{equation}
where the $\langle \cdot \rangle_b$ are the positive functional described in \eqref{eq:bracketsdefinition}, $b=1,\ldots,k$ and their number depends on whether we are considering $f$-coefficients or not. We define $V_{g_i}^{b}(m^2,J)$ as the high-energy part of $g_i$ corresponding to the $k$-functional, following the notation in \cite{Caron-Huot:2020cmc}. In the same way we write the null constraints as
\begin{equation}
0=\sum_{b}\left\langle X_i^{b}(m^2,J) \right\rangle_b\,,
\end{equation}

To get a double-side bound involving the constants $g_i$ and $g_j$ we follow the same strategy in \cite{Caron-Huot:2020cmc}, generalizing it for the case of an arbitrary number of positive functionals $\langle \cdot \rangle_b$, $b=1,\ldots,k$. Let us define
\begin{align}
\begin{split}
\vec V^1 (m^2 ,J) &:=\big(V_{g_i}^{1} (m^2 ,J),V_{g_j}^{1} (m^2 ,J),\vec X^{1}(m^2,J)\big), \\
\vec V^2 (m^2 ,J) &:=\big(V_{g_i}^{2} (m^2 ,J),V_{g_j} ^{2} (m^2 ,J),\vec X^{2}(m^2,J)\big), \\
\vdots & \\
V^k (m^2 ,J) &:=\big(V_{g_i}^{k} (m^2 ,J),V_{g_j} ^{k} (m^2 ,J),\vec X^{k}(m^2,J)\big),
\end{split}
\end{align}
where $\vec X$ represents an arbitrary number of null constraints. With the following algorithm we are able to get a lower and upper bound for $g_j$ with respect to $g_i$, or vice versa.

First of all, we remark that
\begin{equation}
\begin{pmatrix}
\langle \cdot \rangle_1 \\
\langle \cdot \rangle_2 \\
\vdots \\
\langle \cdot \rangle_k
\end{pmatrix}
\cdot 
\begin{pmatrix}
V_1 \\
V_2 \\
\vdots \\
V_k
\end{pmatrix}
=\big(g_i,g_j,\vec 0\,\big).
\end{equation}
To obtain an upper bound, we solve the following optimization problem:
\begin{equation}
\begin{cases}
\text{Min} & B \\
\text{s.t.} & 0 \leqslant (B,-1,\vec c\,)\cdot \vec V^1(m^2,J) \qquad \forall \, m^2\geqslant M^2,\ \forall\,J\in I_1\,,\\
& 0 \leqslant (B,-1,\vec c\,)\cdot \vec V^2(m^2,J) \qquad \forall\, m^2\geqslant M^2,\ \forall\,J\in I_2 \,,\\
&\vdots \\
& 0 \leqslant (B,-1,\vec c\,)\cdot \vec V^k(m^2,J) \qquad \forall \, m^2\geqslant M^2,\ \forall\,J\in I_k\,,
\end{cases}
\label{eq:LowerBound}
\end{equation}
where $c$ is a set of free coefficients which correspond to the number of null constraints used for defining $V$. This relation should hold true for each value of spin in the definition of the respective $\langle\cdots\rangle_b$, $J\in I_b$.. If this is the case, it implies that
\begin{equation}
0\leqslant
\begin{pmatrix}
\langle \cdot \rangle_1 \\
\langle \cdot \rangle_2 \\
\vdots\\
\langle \cdot \rangle_k
\end{pmatrix}
\cdot 
\begin{pmatrix}
(B,-1,\vec c\,)\cdot \vec V^1 \\
(B,-1,\vec c\,)\cdot \vec V^2 \\
\vdots\\
(B,-1,\vec c\,)\cdot \vec V^k
\end{pmatrix}
=Bg_i -g_j\,,
\end{equation}
so that
\begin{equation}
g_j  \leqslant Bg_i\,.
\end{equation}
It is easy to see the fundamental importance of positive functionals to relate the high-energy part with the low one, maintaining the bound from the optimization problem.

We repeat a similar strategy for the lower bound: 
\begin{equation}
\begin{cases}
\text{Max} & A \\
\text{s.t.} & 0 \leqslant (-A,1,\vec c\,)\cdot \vec V^1(m^2,J) \qquad  \forall\, m^2\geqslant M^2,\ \forall\, J\in I_1\,,\\
& 0 \leqslant (-A,1,\vec c\,)\cdot \vec V^2(m^2,J) \qquad \forall \,  m^2\geqslant M^2 ,\ \forall \,J\in I_2\,,\\
&\vdots\\
& 0 \leqslant (-A,1,\vec c\,)\cdot \vec V^k(m^2,J) \qquad \forall \,  m^2\geqslant M^2,\ \forall \,J\in I_k\,,
\end{cases}
\label{eq:UpperBound}
\end{equation}
from which
\begin{equation}
0\leqslant
\begin{pmatrix}
\langle \cdot \rangle_1 \\
\langle \cdot \rangle_2 \\
\vdots\\
\langle \cdot \rangle_k
\end{pmatrix}
\cdot 
\begin{pmatrix}
(-A,1,\vec c\,)\cdot \vec V^1 \\
(-A,1,\vec c\,)\cdot \vec V^2 \\
\vdots\\
(-A,1,\vec c\,)\cdot \vec V^k
\end{pmatrix}
=Ag_i -g_j \, ,
\end{equation}
so that
\begin{equation}
g_j \ \geqslant \ Ag_i\,.
\end{equation}

In particular, we mainly use $g_i= g_2$. In this way we can use positivity of $g_2$,  \eqref{subsec:Bounds4deriv} to write a two-sided bound as 
\begin{equation}
A\leqslant \frac{g_j }{g_2} \leqslant B.
\end{equation}
In other sections, we use two different versions of this optimization problem. One with 3 positive functionals \eqref{eq:bracketsdefinition}, which bounds only $g$-coefficients and another one with 4 functionals \eqref{eq:bracketplusmindefinition}, which bounds $f$- and $g$-coefficients.
We will generally use 
\begin{equation}
\tilde{g}_j:=\frac{g_j}{g_2}.
\end{equation}

\subsubsection{2d bounds}
\label{subsubsec:2d bounds}
We can slightly modify the previous optimization problem to get a bound with respect to two low-energy coefficients. In our case, one of them is always $g_2$ which we take to be positive. Let us define
\begin{align}
\begin{split}
\vec V^1_* (m^2 ,J) &:=\big(V_{g_2}^{1} (m^2 ,J),V_{g_i}^{1} (m^2 ,J),V_{g_j}^{1} (m^2 ,J),\vec X^{1}(m^2,J)\big) \, , \\
\vec V^2_* (m^2 ,J) &:=\big(V_{g_2}^{2} (m^2 ,J),V_{g_i}^{2} (m^2 ,J),V_{g_j} ^{2} (m^2 ,J),\vec X^{2}(m^2,J)\big) \, , \\
\vdots & \\
V^k_* (m^2 ,J) &:=\big(V_{g_2}^{k} (m^2 ,J),V_{g_i}^{k} (m^2 ,J),V_{g_j} ^{k} (m^2 ,J),\vec X^{k}(m^2,J)\big) \, ,
\end{split}
\end{align}
Using the previous results, we have a lower and upper bound of $g_i$ and $g_l$ w.r.t $g_2$. This implies that if we plot $g_i$ and $g_l$, we obtain a rectangular allowed region. We can shrink it solving the following optimization problem
\begin{equation}
\begin{cases}
\text{Min/Max} & (A+B\tilde{g_i}) \, , \\
\text{s.t.} & 0 \leqslant (\pm A,\pm B,\mp 1,\vec c\,)\cdot \vec V^1_*(m^2,J) \qquad \forall J \  m\geqslant M^2,\\
& 0 \leqslant (\pm A,\pm B,\mp 1,\vec c\,)\cdot \vec V^2_*(m^2,J) \qquad \forall J \ m\geqslant M^2 ,\\
&\vdots \\
& 0 \leqslant (\pm A,\pm B,\mp 1,\vec c\,)\cdot \vec V^k_*(m^2,J) \qquad \forall J \ m\geqslant M^2,
\end{cases}
\end{equation}
where we substitute to $\tilde{g_i}$ with values in the finite interval $(\tilde{g}_{i,min},\,\tilde{g}_{i,max})$, where $\tilde{g}_{i,min}$ and $\tilde{g}_{i,max}$ are the values obtained from the previous optimization problem.

Mapping many points in the interval, we are able to generate a convex allowed region like in figure~\ref{fig:g3f4}.

\subsubsection{Implementation in SDPB}
To solve the optimization problems \eqref{eq:UpperBound} and \eqref{eq:LowerBound} we use \texttt{SDPB} \cite{Simmons-Duffin:2015qma,Landry:2019qug}. In order to have the right input for the semidefinite optimization problem:
\begin{itemize}
	\item Normalize all $\vec V^i(m^2,J)$ in order to have $m^2$ only in the numerator. This does not create problems, because it consists in rescaling our system for a positive factor. 
	\item Substitute $m^2 \rightarrow M^2(1+x)$. In this way we have a system of polynomials in $x$.
	\item We consider an appropriate finite set of $J\in I_b\cap \{0,1,\ldots,J_{\mathrm{max}}\}$, adding the constraint with $J \rightarrow \infty$.
\end{itemize}
	 We remark that to obtain a stable value we should consider a $J_{\mathrm{max}}$ depending on the number of null constraints used. In fact, for computations with $\sim 10$ null constraints we need $J_{\mathrm{max}} \sim 40$, meanwhile, for $\sim 100$ null constraints, $J_{\mathrm{max}} \sim 120$ to get a stable result.

\acknowledgments

We thank Andrea Guerrieri, Callum Jones, and Noah Steinberg for useful discussions. This project has received funding from the European Research Council (ERC) under the European Union's Horizon 2020 research and innovation programme (grant agreement no.~758903).

\bibliography{cite.bib}

\bibliographystyle{JHEP.bst}

\end{document}